\documentclass[manuscript]{aastex}
\usepackage{cases}

\newcommand{\fpp}[2]{\frac{\partial #1}{\partial #2}}
\newcommand{\vctr}[1]{\mbox{\boldmath $#1$}}
\newcommand{\mrm}[1]{\mathrm{#1}}

\slugcomment{Not to appear in Nonlearned J., 45.}

\shorttitle{In-Situ Prominence Fomarion Model}
\shortauthors{Kaneko and Yokoyama}

\begin{document}


\title{Numerical Study on In-Situ Prominence Formation by 
Radiative Condensation in the Solar Corona}   

\author{T. Kaneko and T. Yokoyama}
\affil{Department of Earth and Planetary Science, The University of Tokyo,
  7-3-1 Hongo, Bunkyo-ku, Tokyo, 113-0033, Japan}
\email{kaneko@eps.s.u-tokyo.ac.jp}

\begin{abstract}
We propose an in-situ formation model for inverse-polarity solar prominences
and demonstrate it using self-consistent 2.5-dimensional magnetohydrodynamics simulations, 
including thermal conduction along magnetic fields and optically thin radiative cooling. 
The model enables us to form cool dense plasma clouds 
 inside a flux rope by radiative condensation, which is regarded as 
an inverse-polarity prominence. 
Radiative condensation is triggered by changes in the magnetic topology, 
i.e., formation of the flux rope from the sheared arcade field, 
and by thermal imbalance due to the dense plasma trapped inside the flux rope.
The flux rope is created by imposing converging and shearing motion
on the arcade field.
Either when the footpoint motion is in the anti-shearing direction 
or when heating is proportional to local density,
the thermal state inside the flux rope becomes cooling-dominant, leading
to radiative condensation.
By controlling the temperature of condensation, we investigate the 
relationship between the temperature and density of prominences and derive 
a scaling formula for this relationship.
This formula suggests that 
the proposed model reproduces the observed density of prominences, 
which is 10--100 times larger than the coronal density.
Moreover, the time evolution of the extreme ultraviolet emission synthesized 
by combining our simulation results with the response function of 
the Solar Dynamics Observatory Atmospheric Imaging Assembly filters
agrees with the observed temporal and spatial intensity shift among multi-wavelength
during in-situ condensation. 
\end{abstract}


\section{Introduction}
Solar prominences are cool dense plasma clouds that form in the hot tenuous corona. 
Dense plasmas are sustained at the dipped (concave-up) structures 
of coronal arcade fields \citep{Kippenhahn1957,KuperusRaadu1974AA}.
Prominences can be classified into two groups 
--- normal- and inverse-polarity prominences ---
depending on their interior magnetic 
polarity with respect to the overlying arcade field.
In normal-polarity prominences, the magnetic polarity matches
that of the overlying arcade field.
Normal-polarity prominences are typically modeled by the
Kippenhahn--Schl\"uter (KS) model.
In inverse-polarity prominences, the interior magnetic
polarity opposes that of the overlying arcade field. 
This prominence group is described by
the Kuperus--Raadu (KR) model, in which
the lower half of the flux rope indicates the inverse-polarity.
Both normal- and inverse-polarity prominences are observed in the solar atmosphere
\citep{Leroy1984AA}.

An unsolved problem regarding prominences is the origin of their cool dense plasmas.
Though several models have been proposed, the actual mechanism 
by which prominences form remains debatable \citep{Mackay2010SSR}. 
One candidate mechanism for prominence formation is 
radiative condensation \citep{Antiochos1999ApJ}.
Several models for triggering radiative condensation have been proposed
and demonstrated by numerical simulations.
One is the evaporation--condensation process \citep{Mackay2010SSR},
in which evaporated flows from the chromosphere lead to radiative condensation
in the coronal loops.
The mechanism and constraints of this model have been investigated 
using one-dimensional hydrodynamics simulations 
\citep{Karpen2006ApJ,LunaKarpen2012ApJ}
and two-dimensional self-consistent magnetohydrodynamics (MHD) simulations 
\citep{Xia2012ApJ}.
These studies have shown the formation
of normal-polarity prominences categorized by the KS model,
whereas \citet{KeppensXia2014ApJ} showed that the evaporation--condensation model
eventually evolved into the funnel prominence in which the flux rope structure is embedded.

\citet{Choe1992} demonstrated the formation of a normal-polarity prominence
by a different process.
In their two-dimensional MHD simulations, 
the shearing motion at the footpoints of the potential arcade fields 
leads to radiative condensation due to the expansion of the arcade fields.
\citet{Linker2001JGR} demonstrated 
the formation of inverse-polarity prominences. 
In their simulations, the chromospheric plasmas are levitated as a result of
flux rope formation by the magnetic cancellation and condensation occurs.

Condensation models triggered by evaporation or levitation
rely on the injection of chromospheric plasmas.
On the other hand, a recent observation of a prominence formation event showed  
no clear evidence of chromospheric plasma injection, and claimed
the presence of in-situ condensation \citep{Berger2012}.
\citet{XiaKeppens2014} modeled in-situ condensation inside a flux rope 
by three-dimensional simulation,
and showed that the features of prominences are well reproduced.
Because they used the mechanically stable flux rope model 
obtained from the isothermal simulation \citep{XiaKeppensGuo2014ApJ} 
and started their simulation with an ad hoc thermal imbalance,
the process required to trigger the radiative condensation is still unclear. 

In this study, 
we model the process from flux rope formation to radiative condensation
using self-consistent 2.5-dimensional MHD simulations, including thermal conduction 
and radiative cooling. 
By comparing the results of multiple simulations with different settings 
regarding the formation model of flux rope and the coronal heating model, 
we discuss the condition for triggering radiative condensation and
the possible relationship between the temperature and density of the prominence condensations. 

Section 2 introduces our prominence formation model, and 
Section 3 describes the numerical simulation setup.
The simulation results are presented in Section 4. 
The condition for triggering radiative condensation and
a scaling formula from the possible temperature--density relationships
are discussed in Section 5.

\section{Possible processes for triggering in-situ condensation}
\begin{figure}[htbp]
  \begin{center}
    \includegraphics[scale=0.6,bb=0 0 593 200]{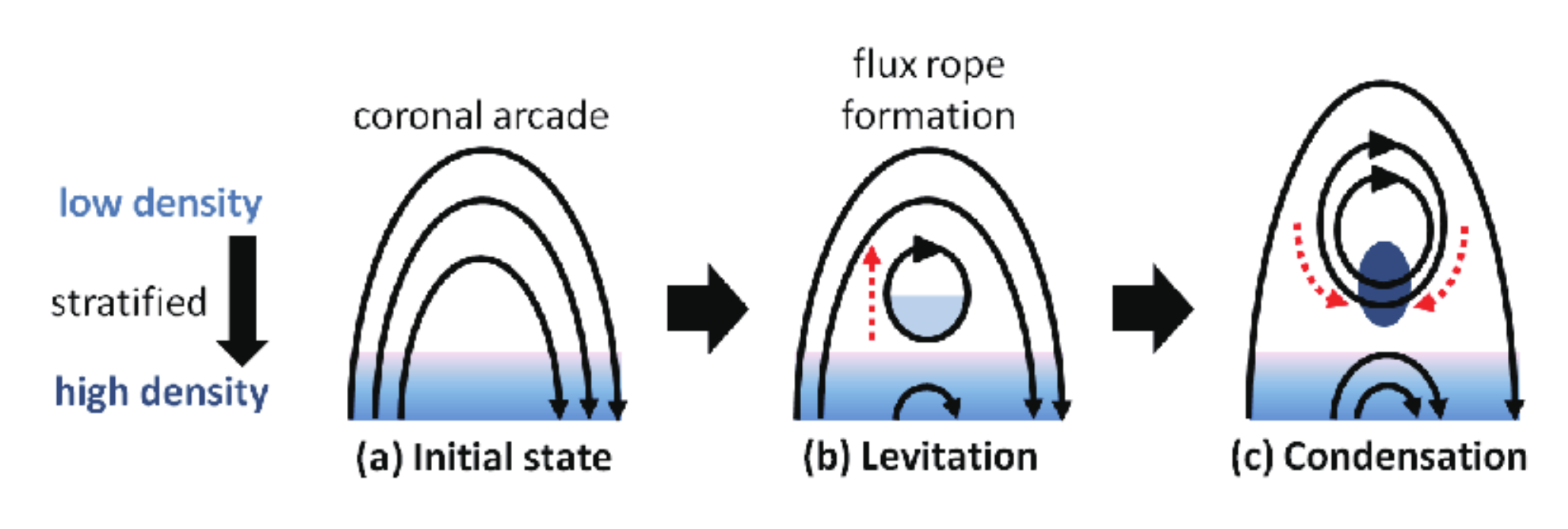}
    \caption{Schematic of a possible process of in-situ radiative condensation}
    \label{levitation}
  \end{center}
\end{figure}

The schematic picture of our model for triggering in-situ radiative condensation is presented 
as follows. Initially, the coronal arcade exists in thermal equilibrium (Fig. \ref{levitation} (a)).
The arcade field is transformed into a flux rope structure by
converging and shearing motion \citep{vanBallegooijen1989AA,MartensZwaan2001ApJ}.
The relatively dense plasmas in the lower corona due to stratification
are trapped inside closed loops of the flux rope, 
where they are elevated into the upper corona (Fig. \ref{levitation} (b)).
The dense plasmas increase the radiative cooling in the flux rope, leading to
thermal imbalance.
Because the closed field lines are disconnected from the exterior region,
the heat loss by radiative cooling inside the flux rope is not conducted outward,
and radiative condensation can be triggered.
(Fig. \ref{levitation} (c)).

\section{Numerical Settings}
To demonstrate the proposed model, we perform 2.5-dimensional MHD simulations, 
including thermal conduction along the magnetic field lines, 
radiative cooling, and gravity in a Cartesian ($x$,$y$) domain.
To investigate the necessary condition for triggering radiative condensation,
we test different direction of the converging and shearing motion 
for the formation of a flux rope and different coronal heating models.

\subsection{Initial and boundary conditions}
The corona model is set up in a rectangular box,
in which the $x$- and $y$-axes are horizontal and vertical, respectively, 
and the $z$-axis is orthogonal to the $x$--$y$ plane.
The corona is stratified under uniform temperature 
($T_{\mrm{cor}}=\mrm{1MK}$) and gravity 
($g_{\mrm{cor}}=\mrm{2.7\times 10^{4}~cm~s^{-2}}$) conditions,
\begin{eqnarray}
  \rho (y)&=&\rho _{\mrm{cor}}\exp \left[ 
-\frac{mg_{\mrm{cor}}}{k_{B}T_{\mrm{cor}}}y\right], \\
  p(y)&=&\frac{k_{B}}{m}\rho (y)T_{\mrm{cor}},
\end{eqnarray}
where $\rho _{\mrm{cor}}=\mrm{3.2 \times 10^{-15}~g~cm^{-3}}$,
$k_{B}$ is Boltzmann's constant, and $m$ is the mean molecular mass. 
We set $m=m_{p}$ with $m_{p}$ being the proton mass.
The force-free arcade field iss described as
\begin{eqnarray}
B_{x}&=&-\left( \frac{2L_{a}}{\pi a}\right)B_{a}\cos \left( \frac{\pi }{2L_{a}}x
		\right)\exp \left[ -\frac{y}{a} \right], \label{ar1}\\
B_{y}&=&B_{a}\sin \left( \frac{\pi }{2L_{a}}x \right)
 \exp \left[ -\frac{y}{a} \right], \label{ar2}\\
B_{z}&=&-\sqrt{1-\left( \frac{2L_{a}}{\pi a} \right)^{2}}B_{a}\cos \left( 
		  \frac{\pi }{2L_{a}}x\right) \exp \left[ -\frac{y}{a}
					      \right], \label{ar3} 
\label{ar3}
\end{eqnarray}
where $B_{a}=3~\mrm{G}$ is the field strength at the footpoint, $L_{a}=12~\mrm{Mm}$ is the width, 
and $a=31~\mrm{Mm}$ is the magnetic scale height of the arcade field. 
Initially, the system exists in mechanical equilibrium.

The left and right boundaries are subjected to 
symmetric (for $\rho ,p,v_{y},B_{y}$) or  
anti-symmetric (for $v_{x},v_{z},B_{x},B_{z}$) boundary conditions. 
A free boundary condition is applied to the top.
In the region below $y=0$, 
the converging and shearing motions are introduced.
We test three types of the footpoint motions.
One is the converging motion without shearing;
the others are converging motion with shearing
that increases the magnetic shear of the arcade field 
or with anti-shearing that decreases the shear of the arcade field.    
In all the cases,
the velocity components $v_{x}$ and $v_{y}$ within this region 
are set as follows,
\begin{eqnarray}
  v_{x} &=& -v_{0}(t)\frac{x}{L_{a}/4},~v_{y}=0,~~(0 \le x<L_{a}/4,~y<0) \label{nc1}\\
  v_{x} &=& -v_{0}(t)\frac{(L_{a}-x)}{3L_{a}/4},~v_{y}=0,~~(L_{a}/4 \le x \le L_{a},~y<0),
  \label{nc2}
\end{eqnarray}
\begin{numcases}{v_{0}(t)=}
  v_{00}, ~~(0<t<t_{1}) \\
  v_{00}(t_{2}-t)/(t_{2}-t_{1}), ~~(t_{1} \le t<t_{2}) \\
  0,~~(t \ge t_{2})
\end{numcases}
where $v_{00}=12~\mrm{km/s}$, $t_{1}=1000~\mrm{s}$, and $t_{2}=1500~\mrm{s}$.
In the case of no shearing motion, $v_{z}=0$.
The shearing and the anti-shearing motions are set as follows,
\begin{eqnarray}
  v_{z} &=& \pm v_{0}(t)\frac{x}{L_{a}/4},~~(0 \le x<L_{a}/4,~y<0)\\
  v_{z} &=& \pm v_{0}(t)\frac{(L_{a}-x)}{3L_{a}/4},~~(L_{a}/4 \le x \le L_{a},~y<0)
\end{eqnarray}
where the plus and minus signs represent the shearing and anti-shearing motions, respectively.
The magnetic fields are computed with the induction equation
by coupling with the given converging and shearing motions.  
The magnetic fields at the bottom boundary are fixed to the initial state.
The gas pressure and density are fixed by assuming hydrostatic equilibrium 
at a constant temperature of $T_{\mrm{cor}}=10^{6} ~\mrm{K}$.

\subsection{Basic equations}
The basic equations are
\begin{equation}
\fpp{\rho }{t}+\vctr{v}\cdot \vctr{\nabla }\rho =-\rho \vctr{\nabla }\cdot \vctr{v}, 
\label{eq_mass_nc}
\end{equation}
\begin{equation}
\fpp{e_\mrm{th}}{t}+\vctr{v}\cdot \vctr{\nabla } e_\mrm{th} =-(e_\mrm{th}+p) 
\vctr{\nabla }\cdot \vctr{v}+\vctr{\nabla }\cdot \left( \kappa T^{5/2} \vctr{bb}\cdot \vctr{\nabla }T\right)-L,
\label{eq_energy_nc}
\end{equation}
\begin{equation}
  e_\mrm{th}=\frac{p}{\gamma -1},
\label{eth}
\end{equation}
\begin{equation}
\fpp{\vctr{v}}{t}+\vctr{v}\cdot \vctr{\nabla v}=-\frac{1}{\rho }\vctr{\nabla }p 
+ \frac{1}{4\pi \rho} \left( \vctr{\nabla} \times \vctr{B}\right) \times \vctr{B}
+ \vctr{g},
\label{eq_momentum_nc}
\end{equation}
\begin{equation}
\fpp{\vctr{B}}{t}=-c\vctr{\nabla }\times \vctr{E},
\label{eq_induction}
\end{equation}
\begin{equation}
\vctr{E}=-\frac{1}{c}\vctr{v}\times \vctr{B}+\frac{4\pi \eta}{c^{2}}\vctr{J}, 
\label{eq_ohm}
\end{equation}
\begin{equation}
  \vctr{J}=\frac{c}{4\pi }\vctr{\nabla }\times \vctr{B},
  \label{eq_current}
\end{equation}
where $\mrm{\kappa =2\times 10^{-6}~erg~cm^{-1}~s^{-1}~K^{-7/2}}$ 
is the coefficient of thermal conduction, $\vctr{b}$ 
is a unit vector along the magnetic field, and
$\vctr{g}=(0,-g_{\mrm{cor}},0)$ is the gravitational acceleration
and $\eta $ is the magnetic diffusion rate.
The temperature is computed by the following equation of state,
\begin{equation}
  T=\frac{m}{k_{B}}\frac{p}{\rho }.
\end{equation}
For fast reconnection, 
we adopt the following form of the anomalous resistivity
\citep[e.g.][]{YokoyamaShibata1994ApJ},
\begin{numcases}{\eta =}
  0, ~~~~~~~~~~~~~~~~~~~\left(J < J_{c}\right)\\
  \eta _{0}\left(J/J_{c}-1\right)^{2}, ~~\left(J \ge J_{c}\right)
\end{numcases} 
where $\eta _{0}=\mrm{3.6\times 10^{13}~cm^{2}~s^{-1}}$ and
$J_{c}=\mrm{25~dyn^{1/2}~cm~s^{-1}}$. 
We restrict $\eta $ to $\eta _{\mrm{max}}=\mrm{18.0\times 10^{13}~cm^{2}~s^{-1}}$.
$L$ in Eq. (\ref{eq_energy_nc}) represents the net cooling term, which is expressed as
\begin{equation}
  L=n^{2}\Lambda (T)-H-\eta J^{2},
\end{equation}
where $n=\rho /m$ is the number density, 
$\Lambda (T)$ is the radiative loss function of  
optically thin plasma, $H$ is the background heating rate,
and $\eta J^{2}$ represents ohmic heating.
We adopt a simplified radiative loss function in Fig. \ref{loss} (a) \citep{Hildner1974SoPh}
and simulate two different coronal heating models. 
In one model, the heating rate $H$ depends on the local magnetic energy density 
(magnetic pressure) at temperatures above the cutoff temperature 
$T_{\mrm{c}}$ and is balanced out by the cooling rate when it falls below $T_{\mrm{c}}$,
\begin{numcases}{H=}
  \alpha _{m}P_{m},~~\left(T \geq T_{\mrm{c}}\right) \label{ht_a_1}\\
  n^{2}\Lambda (T), ~~\left( T < T_{\mrm{c}} \right) \label{ht_a_2}
\end{numcases}
Initial thermal equilibrium required that
\begin{equation}
  \alpha _{m}=\frac{n_\mrm{cor}^{2}\Lambda (T_\mrm{cor})}{B_{a}^{2}/8\pi }
  \exp \left[-2\left(\frac{mg_\mrm{cor}}{k_{B}T_\mrm{cor}}-\frac{1}{a} \right)y\right],
\end{equation}
where $n_\mrm{cor}=\rho _\mrm{cor}/m=2.0\times 10^{9}~\mrm{cm^{-3}}$. 
Selecting $a=k_{B}T_\mrm{cor}/(mg_\mrm{cor})$, we obtain 
$\alpha _{m}=6.2\times 10^{-4}~\mrm{cm^{3}~s^{-1}}$ (i.e., a constant).
Equation (\ref{ht_a_2})
prevents the temperature from decreasing below $T_{\mrm{c}}$.
In the other model, the heating rate $H$ depends on the local density and height 
at temperatures above $T_{\mrm{c}}$ 
and balances with the cooling rate below the cutoff temperature $T_{\mrm{c}}$.
\begin{numcases}{H=}
  \alpha _{d}(y) n,~~\left(T \geq T_{\mrm{c}}\right) \label{ht_b_1}\\
  n^{2}\Lambda (T). ~~\left( T < T_{\mrm{c}} \right) \label{ht_b_2}
\end{numcases} 
For the initial thermal equilibrium, we have 
$\alpha _{d}(y)=n_\mrm{cor} \Lambda (T_\mrm{cor})
\exp [-mg_\mrm{cor}y/(k_{B}T_\mrm{cor})]$ where 
$n_\mrm{cor} \Lambda (T_\mrm{cor})=1.1 \times 10^{-13}~\mrm{erg~s^{-1}}$. 
Note that ohmic heating is always much smaller than background heating
in these settings.
 $T_\mrm{c}$ is one of the model parameters. 
For each heating and shearing model, we
vary $T_\mrm{c}$ as $0.3$ MK, $0.4$ MK, and $0.5$ MK.
The investigated cases are summarized in Table \ref{table}.

\subsection{Numerical scheme and grid size}
The numerical scheme is based on the CIP-MOCCT method \citep{Kudoh1999}.
To prevent numerical oscillations in the advection term 
(left-hand side of Eqs. (\ref{eq_mass_nc})--(\ref{eq_momentum_nc})),
we apply the rational CIP scheme \citep{Xiao1996}, 
in which the rational function replaces the cubic function as an interpolation function of CIP.
The thermal conduction, radiative cooling, and heating terms are included 
as source terms  in non-advection phases of CIP method.
Thermal conduction is  explicitly solved 
by a  second-order accurate slope-limiting method {\citep{SharmaHammett2007JCoPh}.

The required grid size for preventing numerical thermal instability can be 
estimated by the formula presented in \cite{KoyamaInutsuka2004}.
When applying this formula to our simulation, we assume
that the density of the condensations is 10 times the coronal density,
\begin{equation}
  \Delta <\frac{\lambda _{F}}{3}=\frac{1}{3}
  \sqrt{\frac{\kappa T_\mrm{c}^{7/2}}{n _\mrm{c}^{2}
      \Lambda (T_\mrm{c})}}, \label{ki}
  \label{rgrids}
\end{equation}
where $\Delta $ is the grid size, $\lambda _{F}$ is the Field length, 
$T_\mrm{c}$ is the cutoff temperature 
(also assume to be the temperature of condensations), 
and $n _\mrm{c}=10n _\mrm{cor}$ is the assumed density of condensations.
Figure \ref{loss} (b) shows the required grid size against temperature. 
Under the constraint of Eq. (\ref{ki}), the grid size is determined to be  
$\Delta =$ 120 km, 60 km, and 30 km in the cases of   
$T_{\mrm{c}}=$0.5 MK, 0.4 MK, and 0.3 MK, respectively.
The horizontal grid size $\Delta x$ is uniform. 
The vertical grid size $\Delta y$ is uniform up to $y<24~\mrm{Mm}$ 
and increases to $\Delta y =\mrm{1.8\times 10^{3}~km}$ 
at $y>24~\mrm{Mm}$ to ensure a sufficiently distant upper boundary.

\begin{figure}[htbp]
  \begin{center}
    \includegraphics[bb=0 0 425 170]{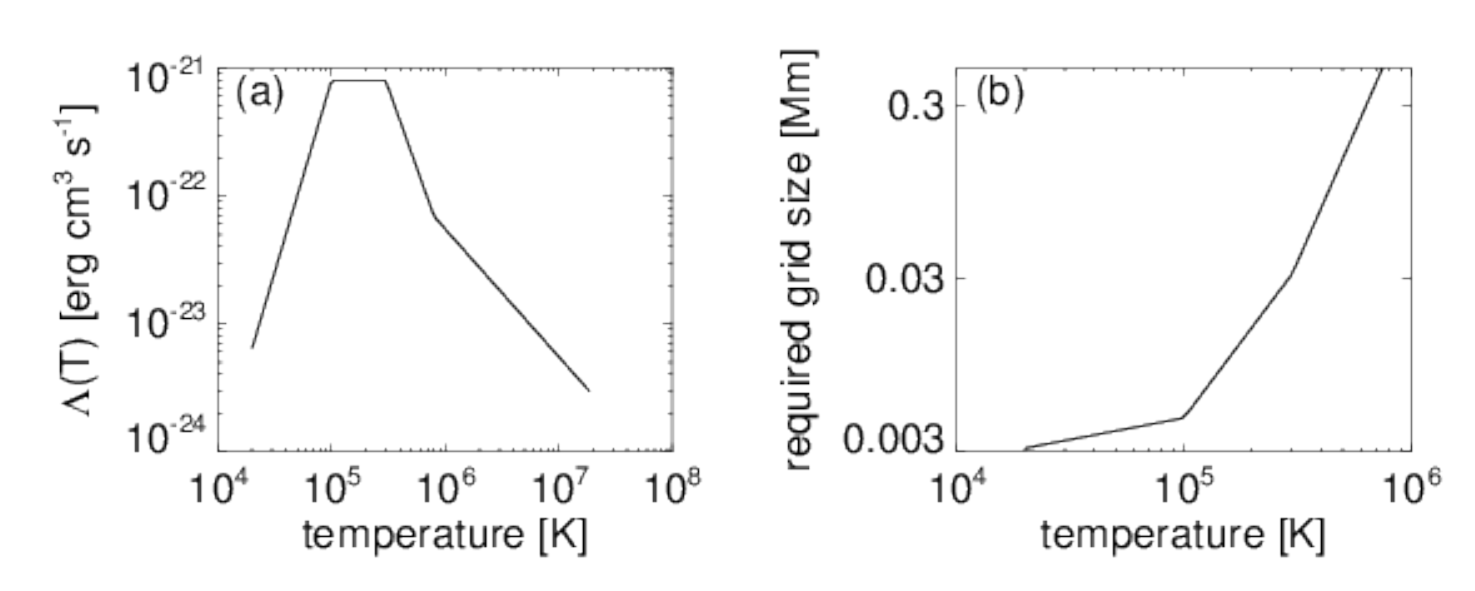}
    \caption{Simplified radiative cooling functions and required grid sizes}
   \label{loss}
  \end{center}
\end{figure}

\section{Results} \label{bozomath}
The right-most column of Table \ref{table} shows our results in the presence of 
the radiative condensation. 
We find that the necessary condition for the radiative condensation is either anti-shearing
motion or the heating proportional to the local density.
Panels (a)--(d) of Fig. \ref{snap_c_pm} are snapshots of the time evolution and
Fig. \ref{snap_c_pm2} is the final state in case M1,
which is a the typical case for the radiative condensation.
The initial state exists in mechanical and thermal equilibrium (Fig. \ref{snap_c_pm} (a)). 
Converging motion triggers reconnection above the PIL at $x=0$, 
and a flux rope is formed (Fig. \ref{snap_c_pm} (b)). 
As the reconnection proceeds, the relatively dense plasmas in the lower corona 
are trapped inside the flux rope and lifted to the upper corona. 
The dense plasmas enhances radiative cooling
in the flux rope, which overwhelms the background heating and leads to thermal imbalance
(Fig. \ref{snap_c_pm} (c)).
Because thermal conduction works only along closed magnetic field lines 
of the flux rope, it can not compensate for the thermal imbalance.
Consequently, radiative condensation is triggered, and the cool dense plasmas 
accumulates in the lower part of the flux rope (Fig. \ref{snap_c_pm} (d)).  
Figure \ref{snap_c_pm2} (b) shows
profiles of temperature, number density, and horizontal component of the 
magnetic field along the $y$-axis in the quasi--static state (Fig. \ref{snap_c_pm2} (a)).
The temperature of the prominence 
corresponding to $T_\mrm{c}$ is approximately
$0.3~\mrm{MK}$ in this case.
The heating settings (Eqs. (\ref{ht_a_2}) and (\ref{ht_b_2})) maintains
the temperature of the cool dense plasmas at around $T_\mrm{c}$.  
According to Fig. \ref{snap_c_pm2} (b), the magnetic polarity of the
cool dense region opposes that of the initial arcade field.
Figures \ref{snap_c_pm2} (a) and (b) clearly show that
radiative condensation eventually establishes inverse-polarity prominence
in the flux rope. 
Also, the high-density region (warm color region in the density plot 
of Fig. \ref{snap_c_pm2} (a)) resembles the tower prominence 
\citep{Berger2012}, 
and the low-density region surrounding the prominence (dark region in the same plot)
can be regarded as the coronal cavity.
Figure \ref{vforce} shows each force at a time of $=$4250s.
The mechanical quasi-static state is realized in the vertical direction.
The upward magnetic tension force is balanced with the downward pressure 
gradient force and gravity, i.e., the prominence mass is supported 
mainly by the magnetic tension. This force balance system is also found
in the parametric study by adiabatic MHD simulations of \citet{Hillier2013ApJ}. 
Our result mostly corresponds to their case (e) of model 1 regarding the force balance, 
whereas the prominence density and plasma $\beta $ in our simulation 
are lower and higher than those in \citet{Hillier2013ApJ}, respectively.
They revealed that the magnetic tension force is given 
by the vertical stretching of the flux rope, 
and the curvature of the field lines inside the prominence can be smaller
because the sufficiently large magnetic tension comes from 
the larger field strength by compression of the prominence mass.
Our simulation realizes these features of prominence support as shown 
in Fig. \ref{snap_c_pm2} (a).
Figures \ref{snap_c_llr} shows the time evolution of case M5, 
which is a typical case with no radiative condensation. 
The flux rope is heated-up rather than cooled-down in this case.

\begin{table}[htbp]
  \begin{center}
    \begin{tabular}{ccccc} \hline
      case & Heating & Shearing & $T_{c}$ (MK) &Condensation \\ \hline \hline
      M1 &                    & $-$ & 0.3 & Yes \\ 
      M2 &                    & $-$ & 0.4 & Yes \\ 
      M3 & $H=\alpha _{m}P_{m}$ & $-$ & 0.5 & Yes \\ 
      M4 &                    & 0 &  0.5 & No \\ 
      M5 &                    & $+$ & 0.5 & No \\ \hline 
      D1 &                    & $-$ & 0.3 & Yes \\ 
      D2 &                    & $-$ & 0.4 & Yes \\ 
      D3 & $H=\alpha _{d}\rho $ & $-$ & 0.5 & Yes \\ 
      D4 &                    & 0 & 0.5 & Yes \\ 
      D5 &                    & $+$ & 0.5 & Yes \\ \hline 
    \end{tabular}
    \caption{The presence of radiative condensation in each case.
      The second column shows the heating model.
      The third column shows the shearing model:
      plus sign ($+$), minus sign ($-$), and 0 represent the
      shearing, anti-shearing, and no shearing cases, respectively.
      The fourth column shows the cutoff temperature.
      The fifth column shows the presence of condensation.}
    \label{table}
  \end{center}
\end{table}

\begin{figure}[htbp]
  \begin{tabular}{cc}
    \begin{minipage}{0.5\hsize}
      \includegraphics[scale=0.28,bb=0 0 708 708]{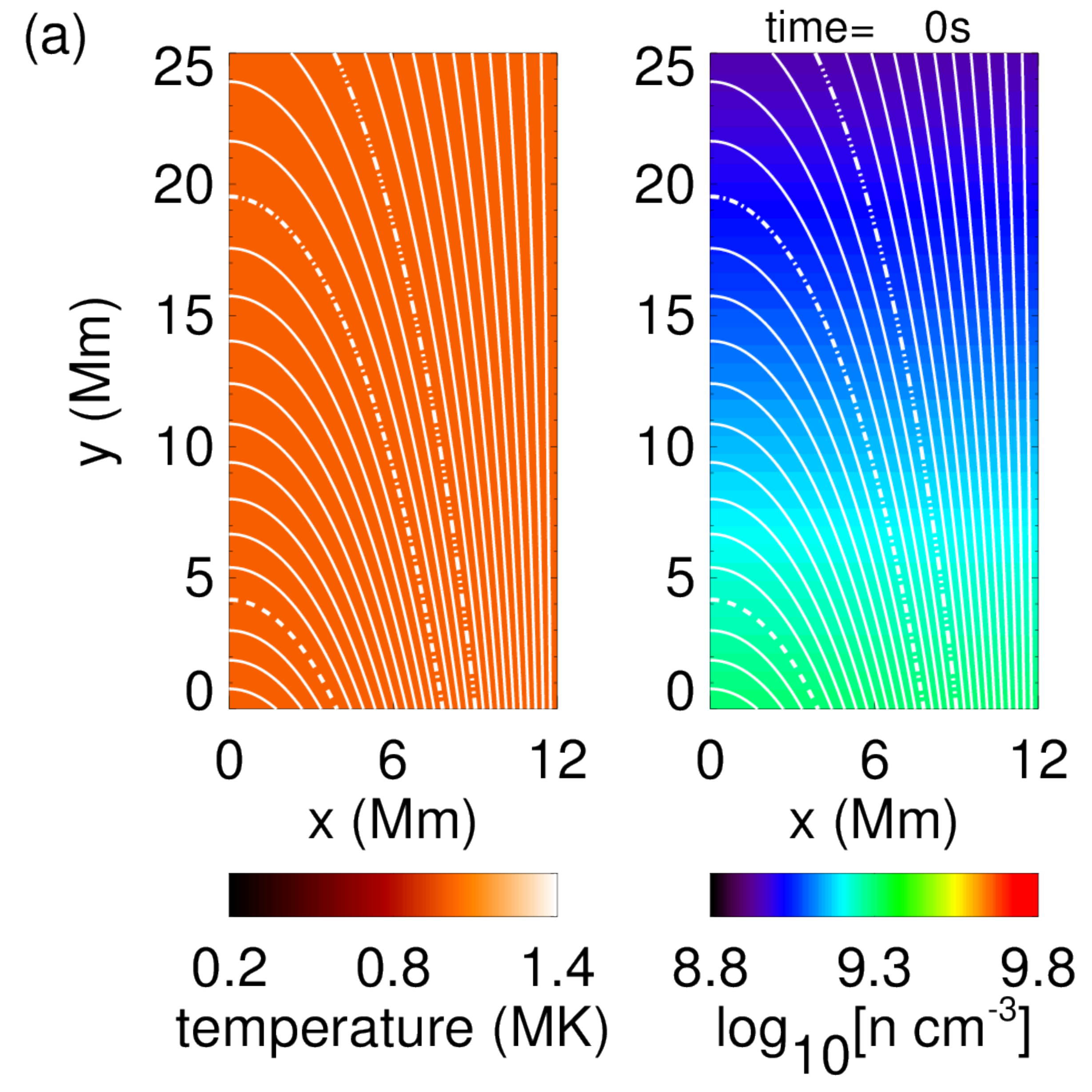}
    \end{minipage}
    \begin{minipage}{0.5\hsize}
      \includegraphics[scale=0.28,bb=0 0 708 708]{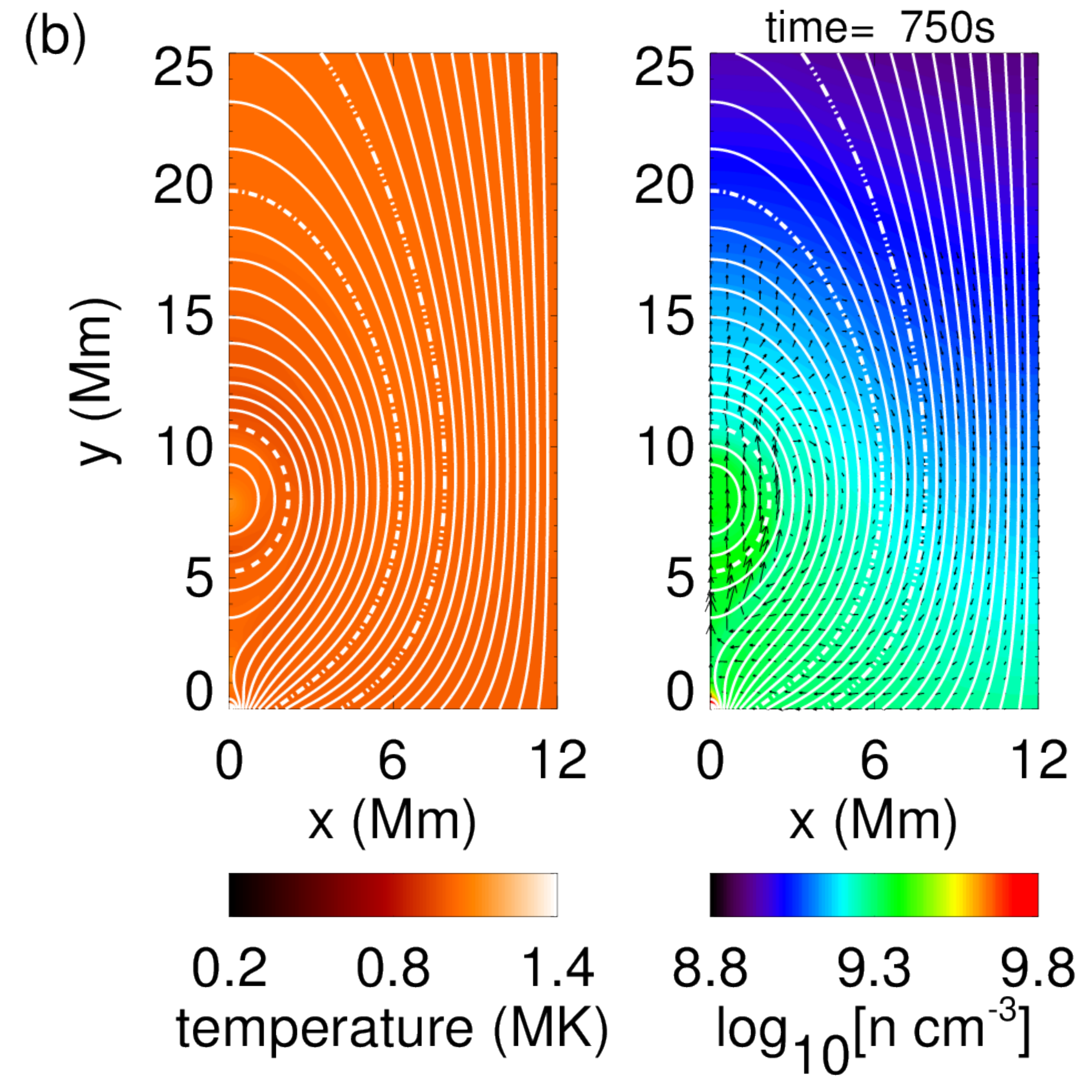}
    \end{minipage}
    \\
    \begin{minipage}{0.5\hsize}
      \includegraphics[scale=0.28,bb=0 0 708 708]{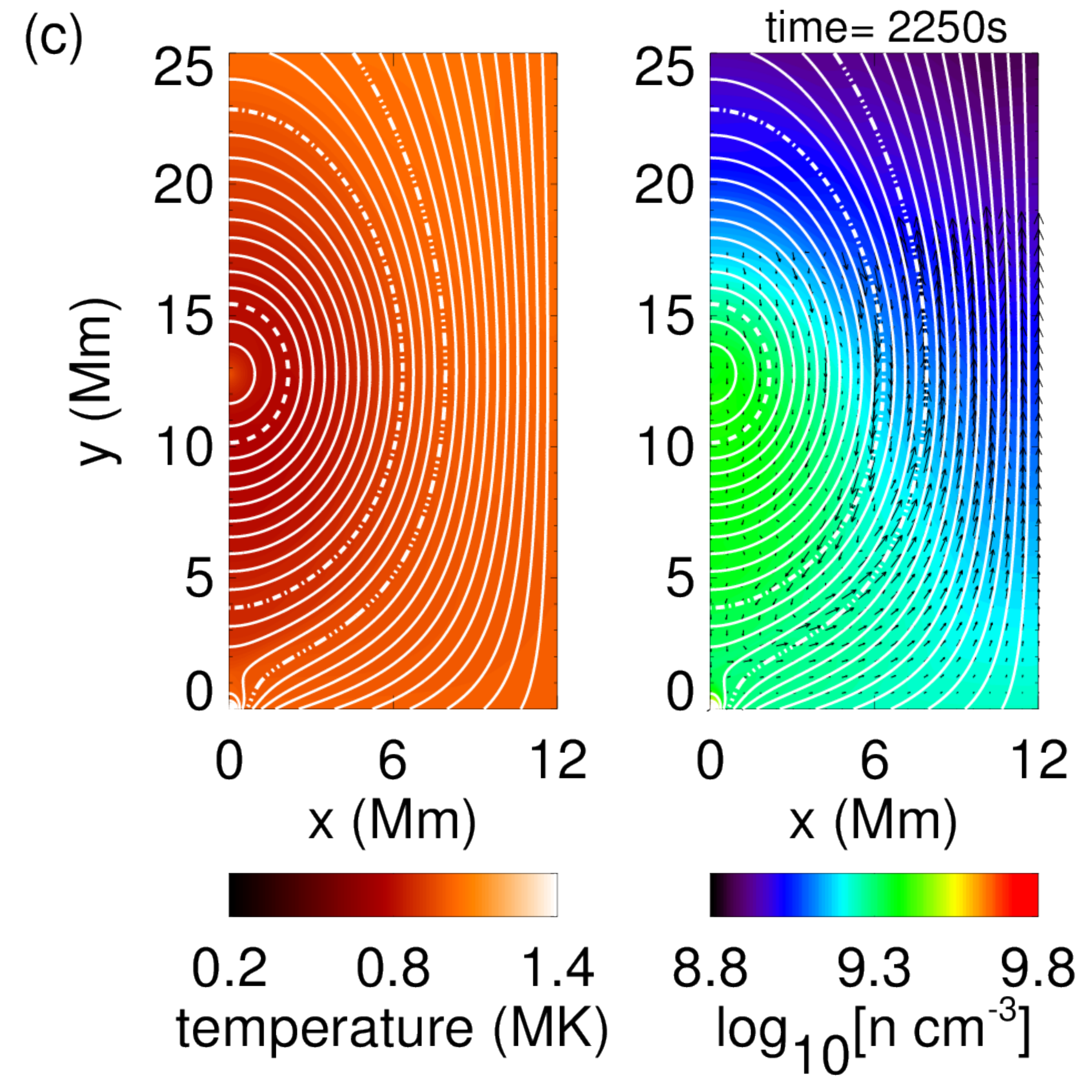}
    \end{minipage}
    \begin{minipage}{0.5\hsize}
      \includegraphics[scale=0.28,bb=0 0 708 708]{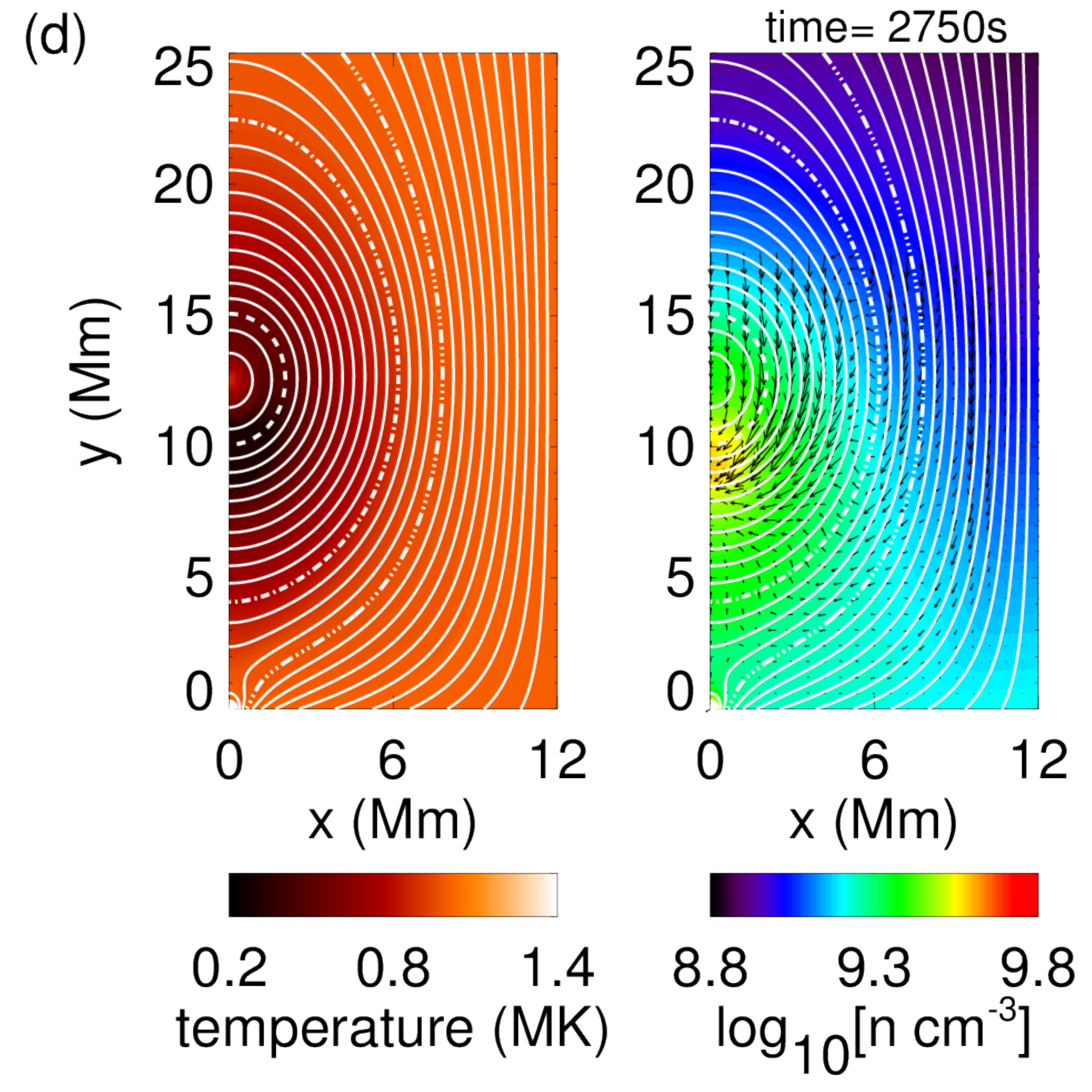}
    \end{minipage}
  \end{tabular}
  \caption{Panels (a)--(d) are snapshots of the time evolution in case M1.
    Colors represent temperature and density,
    white lines represent magnetic field lines, 
    and arrows represent the velocity fields.}
  \label{snap_c_pm}
\end{figure}

\begin{figure}[htbp]
  \begin{tabular}{cc}
    \begin{minipage}{0.5\hsize}
      \includegraphics[scale=0.28,bb=0 0 708 708]{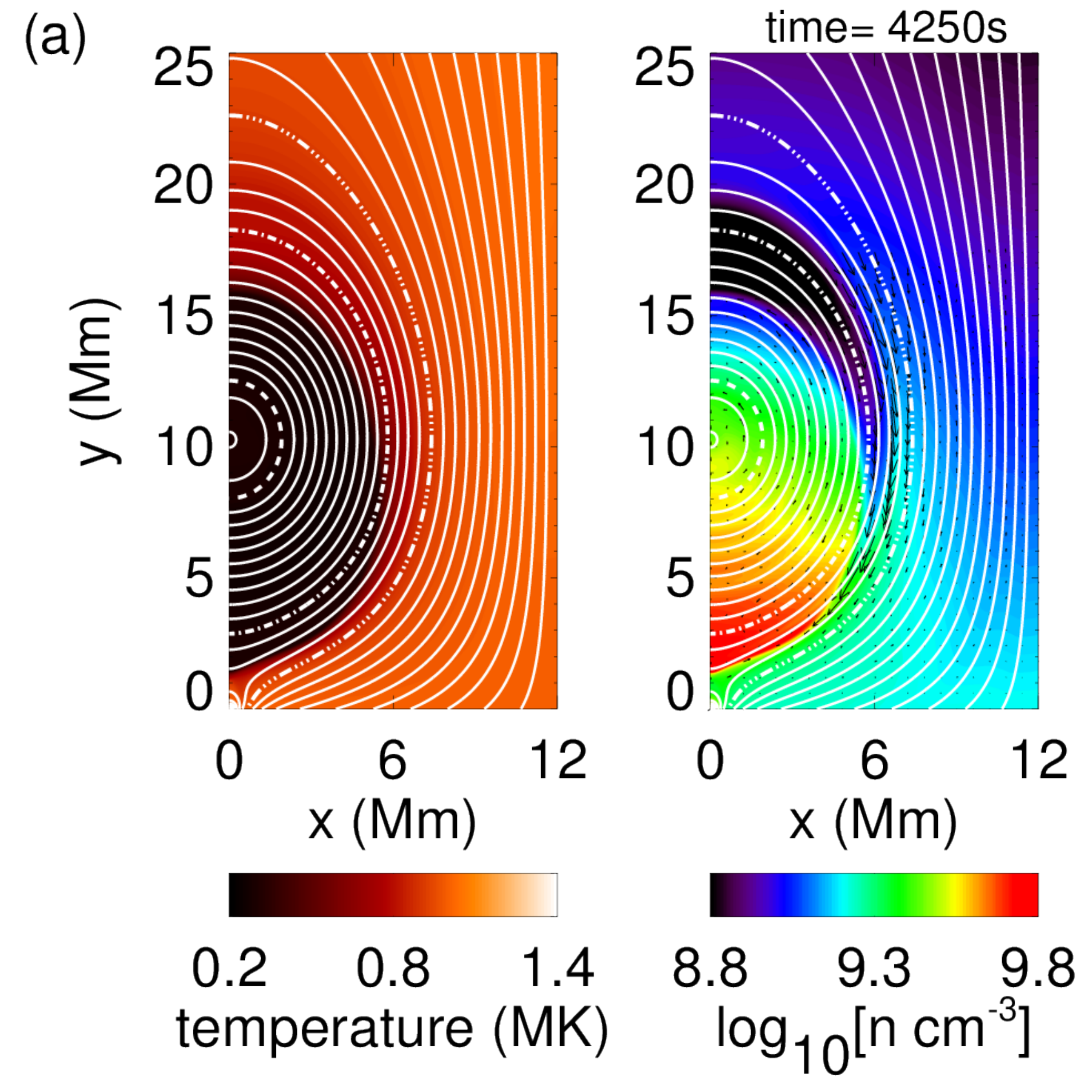}
    \end{minipage}
    \begin{minipage}{0.5\hsize}
      \includegraphics[scale=0.28,bb=0 0 708 708]{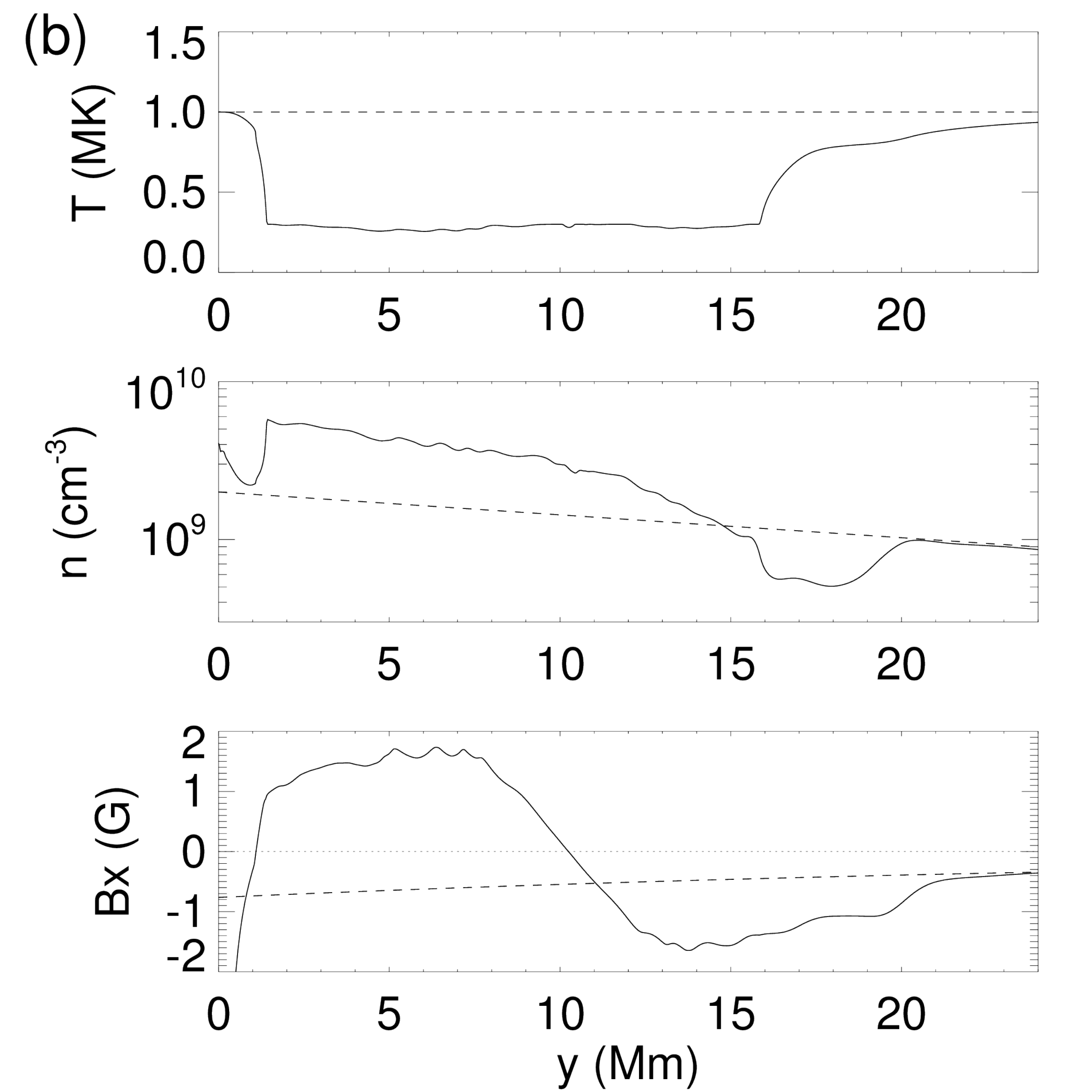}
    \end{minipage}
  \end{tabular}
  \caption{Panel (a) is the final state in case M1.
    Colors, white lines and arrows represent the same quantities as in Fig.\ref{snap_c_pm}. 
    Panel (b) shows the profiles of temperature, number density, and 
    $B_{x}$ along the $y$-axis. 
    The solid and dashed lines are the profiles in the final 
    and initial states, respectively.}
  \label{snap_c_pm2}
\end{figure}

\begin{figure}[htbp]
  \begin{center}
    \includegraphics[scale=0.6,bb=0 0 566 425]{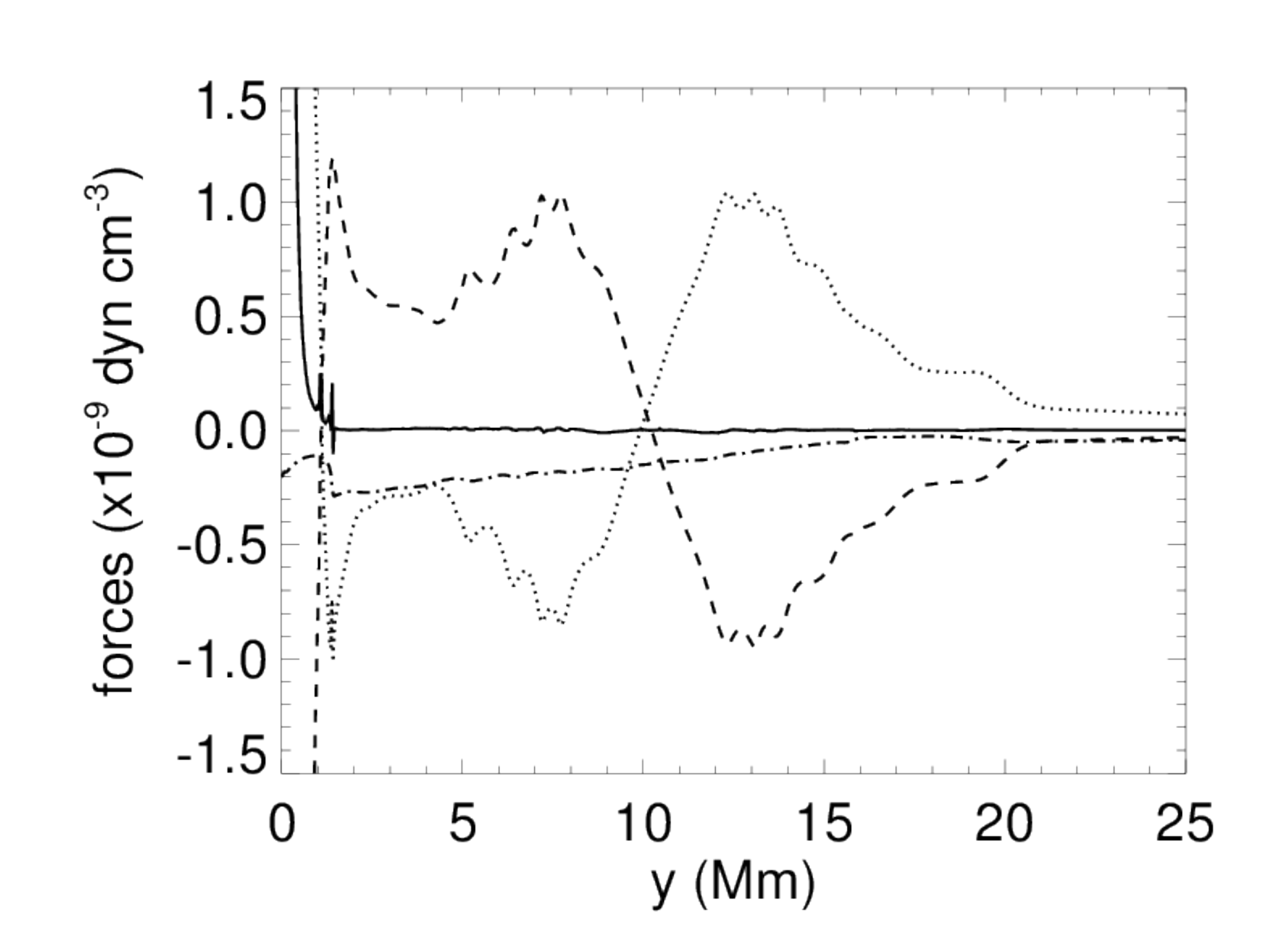}
    \caption{The vertical components of the forces along the $y$-axis. 
      The dotted line, dashed line, and dot-dashed line
      represent the pressure gradient (the sum of the gas and magnetic pressure gradients), 
      magnetic tension force, and gravity, respectively. 
      The solid line represents sum of all the forces.}
    \label{vforce}
  \end{center}
\end{figure}

\begin{figure}[htbp]
  \begin{tabular}{cc}
    \begin{minipage}{0.5\hsize}
      \includegraphics[scale=0.28,bb=0 0 708 708]{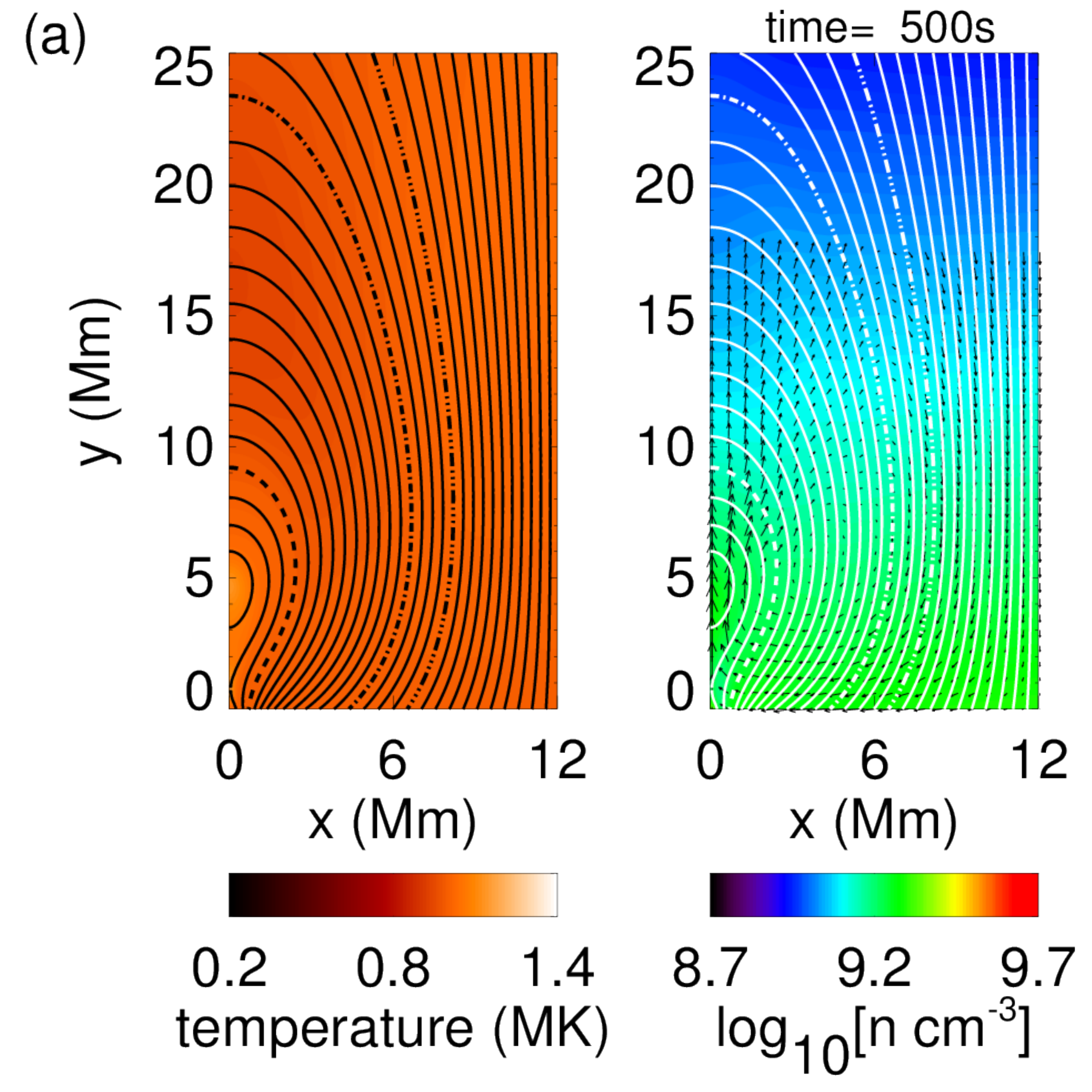}
    \end{minipage}
    \begin{minipage}{0.5\hsize}
      \includegraphics[scale=0.28,bb=0 0 708 708]{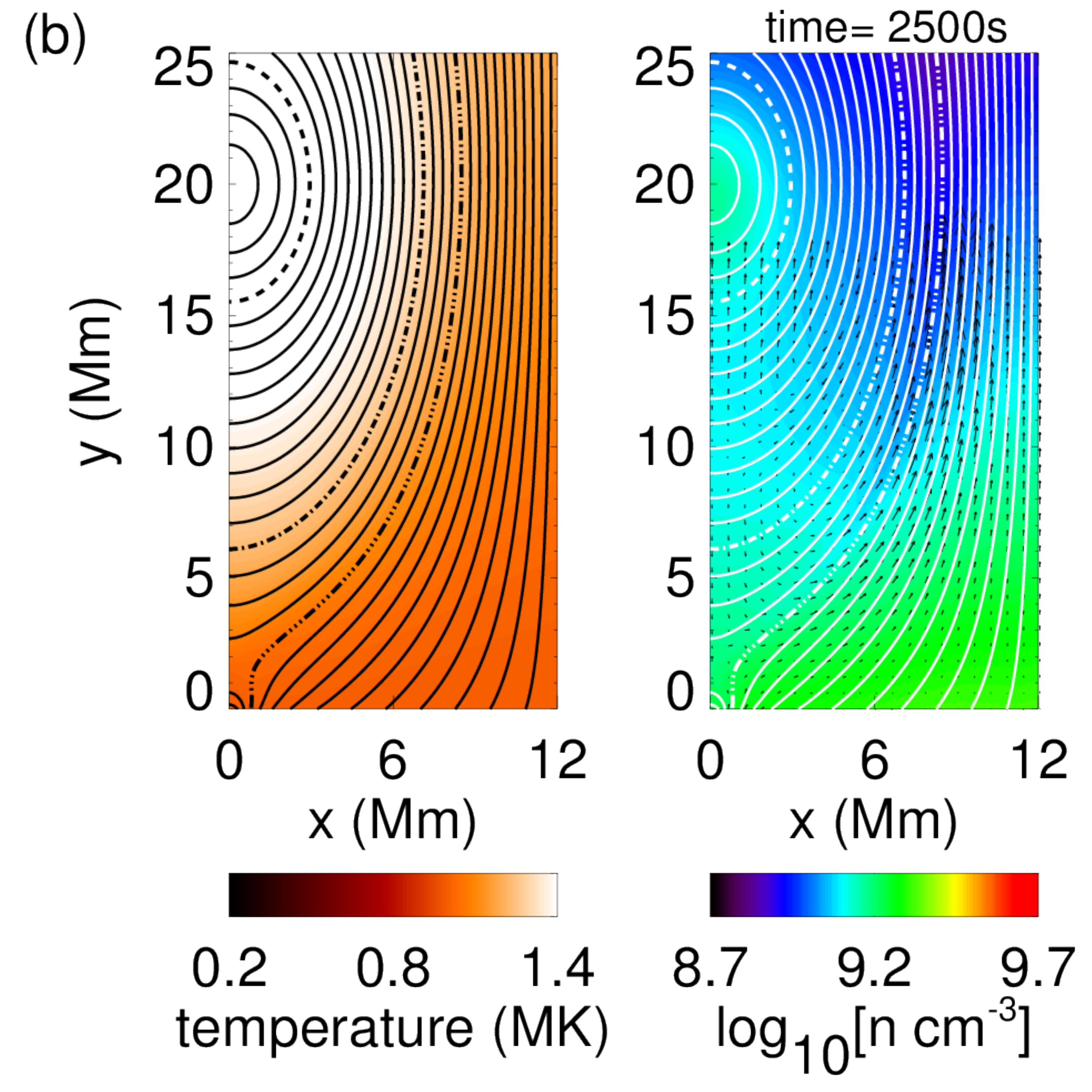}
    \end{minipage}
  \end{tabular}
  \caption{Panels (a)--(b) are snapshots of the time evolution in case M5.
    Colors and arrows denote the same quantities as those in Fig.\ref{snap_c_pm}. 
    Black lines represent magnetic field lines.}
  \label{snap_c_llr}
\end{figure}

\section{Discussion}
As shown in Table \ref{table}, the radiative condensation occurs
either when the anti-shearing motion is imposed or when the heating model 
is proportional to the density, i.e., $H=\alpha _{d}\rho $.
The radiative condensation is triggered by the 
cooling-dominant thermal imbalance in the flux rope composed of the thermally 
isolated closed loops.  
The interior of the flux rope is relatively dense because of the converging
motion as well as the initial stratification.
Because the radiative cooling inside the flux rope is enhanced by the dense plasma, 
the cooling-dominant thermal imbalance is reasonable
if the heating changes only slightly.
Because we set the heating depending on the magnetic pressure 
(cases M1--M5, hereafter called M-cases)
and local density (cases D1--D5, hereafter called D-cases),
the change in heating rate is not negligible.
In some cases (M5 and M6), a heating-dominant thermal imbalance is 
created and no condensations occur.
In the following subsections, we discuss the mechanism for creating the 
cooling/heating-dominant thermal imbalance in the M- and D-cases.
Then, our model is compared with previous theoretical studies and observations.

\subsection{Thermal imbalance in M-cases}
In M-cases where the heating rate is proportional to the magnetic pressure,  
the anti-shearing and shearing motion causes a larger 
discrepancy between the cooling and heating rates, establishing the cooling- and 
heating-dominant thermal imbalances, respectively.
Figure \ref{clht} shows the cooling and heating rates along the $y$-axis at time=$500 \mrm{s}$
in case M3--M5.
In case M3 (anti-shearing case), the cooling rate is the highest of the three cases,
whereas the heating rate is the lowest. Case M5 (shearing case) has the opposite
tendency, where the cooling rate is the lowest and the heating rate is the highest. 
We discuss how these tendencies are established from a perspective of the
evolution of the magnetic field and its influence on the cooling and heating rates.

In the anti-shearing case, the flux rope is pushed upward by the reconnection and
confined by the downward magnetic tension of the overlying arcade field. 
Figure \ref{tens} shows the vertical tension force along the $y$-axis at time=$500~\mrm{s}$
in case M3 (anti-shearing), M4 (no shearing), and M5 (shearing).
The downward magnetic tension force in the anti-shearing case 
is larger than those in the shearing and no shearing cases. 
The reason the anti-shearing case has the larger downward tension force 
is as follows:
the shear angle is related to the height of the arcade field as 
\begin{equation}
  a=\frac{L_{a}}{\pi \cos \theta},
  \label{shear_height}
\end{equation}
where $a$ is the arcade height, $\theta = \arctan (B_{z}/B_{x})$ 
is the shear angle against the positive $x$-axis in the $x$$z$-plane, 
$L_{a}$ is the arcade width, 
and we assume a linear force-free arcade field for simplicity. 
Equation (\ref{shear_height}) indicates that the reduction of magnetic shear by 
the anti-shearing motion makes the arcade field shorter.
Due to the shrink of the arcade field, the flux rope in the anti-shearing case experiences
a larger downward magnetic tension force.

The flux rope in the anti-shearing case is pinched by the overlying arcade field 
and pushed by the upward magnetic tension of the reconnected loops, leading to a 
density enhancement by compression.
Figure \ref{dena} shows the time evolution of area and average density
inside the dash-dotted line in Fig. \ref{snap_c_llr}. 
The average density is computed by dividing the total mass inside the dash-dotted line
by the corresponding area. The total mass is almost the same between these three cases
and is conserved after the flux rope is detached from the bottom boundary. 
In Fig. \ref{dena} (b), the anti-shearing case shows the enhancement of the
average density. This result indicates that the density enhancement inside the flux rope in the 
anti-shearing case is due to the compression by the magnetic tension
as well as the converging motion.

The anti-shearing motion lead to 
the higher cooling rate as well as the lower heating rate.
The anti-shearing motion reduces the magnetic component parallel to the PIL, 
corresponding to $B_{z}$ in our simulations, and decreases the magnetic pressure,
leading to the lowest heating rate.
Thus, the interior of the flux rope in case M3 suffers from unstoppable cooling
most efficiently. 
In case M5 (shearing case), on the other hand, the heating rate increases as 
the magnetic pressure increases, and 
the enhancement of the cooling rate is too small to overwhelm the heating rate.
Eventually, the shearing motion is likely to establish 
the heating-dominant thermal imbalance, resulting only in the formation of hot flux rope.
These results suggest that when the coronal heating depends on the magnetic energy density,
the continuous shearing motion can not trigger the radiative condensation, 
whereas the temporary anti-shearing motion can indeed trigger it, 
leading to the prominence formation.

\subsection{Thermal imbalance in D-cases}
From the results of the D-cases in which the heating is 
proportional to the density, 
it is found that the cooling-dominant thermal imbalance 
can be achieved only by the difference between the heating and cooling rates 
in the dependence on the density.
Figure \ref{clht_ro} shows the cooling and the heating rates in the cases D3--D5. 
The cooling rate in case D3 (anti-shearing case) is highest 
because the density enhancement is largest in the anti-shearing case,
as discussed in the previous subsection.
Unlike case M3, however, the heating rate in case D3 is also highest.
The combination of the higher cooling rate and the higher heating rate
can be a disadvantage to enhancing the thermal imbalance.
What actually sets the cooling-dominant thermal imbalance in D-cases is the
different density dependence between the heating ($H\propto \rho  $) 
and cooling terms ($R\propto \rho ^{2}$).
The increment of the heating rate against the density increase
is always smaller than that of the cooling rate.
Consequently, the cooling overwhelms the heating, 
and the radiative condensation is 
triggered in all D-cases.
These results suggest that if the coronal heating is only proportional 
to the density, the flux ropes in the corona are likely to cool down. 
The exact modeling of the coronal heating is required to confirm this mechanism. 

\subsection{Comparison with other theoretical studies}
In evaporation--condensation models \citep{Antiochos1999ApJ,Karpen2006ApJ}, 
radiative condensation is triggered by the transport of dense plasma
from lower altitudes. Our model is consistent with this mechanism, 
but it adopts a different approach to suppress the thermal conduction effects. 
In the evaporation--condensation model,
thermal conduction effects are suppressed by preparing long magnetic field lines,
whereas in our model, they are suppressed by closed loop formations.
In other words, our model altered the magnetic field topology.
Both our study and that of \citet{Choe1992} assume that photospheric motion triggers 
the prominence formation. 
However, the direction of motion differs between the studies. In \citet{Choe1992}, 
radiative condensation arises by shearing motion along the PIL 
and subsequent expansion of the arcade field.
In our study, converging motion towards the PIL
and subsequent formation of closed magnetic fields (flux rope) are essential precursors
of radiative condensation.
Moreover, whereas their models reproduce normal-polarity prominences, 
our models form inverse-polarity prominences.
The formation of the inverse-polarity prominence is found both in
the results of our model and that of \citet{Linker2001JGR}.
The difference is that our model does not require chromospheric plasma
injection to trigger the radiative condensation;
hence, our model reproduce the in-situ condensation.
\cite{XiaKeppens2014} performed 
three-dimensional simulation of the in-situ condensation inside the flux rope. 
The flux rope system in \cite{XiaKeppens2014} 
was created by the converging and shearing motion 
in the isothermal simulation of \cite{XiaKeppensGuo2014ApJ}.
Because the strategy used to establish cooling-dominant thermal imbalance in 
\cite{XiaKeppens2014} was the parameterized heating,
it was still unclear why the thermal imbalance was created inside the flux rope.
We ensure that the cooling-dominant thermal imbalance can be 
created inside the flux rope by two different mechanisms:
the anti-shearing motion 
or the different dependence on the local density
between the heating and cooling rates.
We find that the direction of the shear motion strongly affected the thermal processes 
when the heating depends on the magnetic energy density,
and it controls the presence of in-situ condensation.
When the heating is proportional to the density, on the other hand,
the cooling-dominant thermal imbalance is always established
by the difference between the cooling and heating terms 
in the dependence on density, leading to radiative condensation.

The formation mechanism through the anti-shearing motion 
smoothly connects to some of the eruptive models. 
These include instability models such as kink and torus instability
\citep{Kliem2004A&A,Torok2005ApJ,Kliem2006PhRvL,FanGibson2007ApJ}.
The kink instability is likely to be triggered because
the anti-shearing motion increases the twist of the flux rope. 
The torus instability also fit because
the reduction of shear can alleviate the critical height of torus instability. 
The other model that fit with our mechanism is the reversed shear model of \citet{Kusano2004ApJ}.
We stopped the anti-shearing motion before the shear reversal. 
If we continued to impose the anti-shearing motion, the flux rope formation, 
radiative condensation, and eruption would be successively occurred 
even without the converging motion. 
Both the shearing and anti-shearing motions
have been considered as the possible 
process of eruptions \citep{Kusano2002ApJ,Amari2003ApJ}.
Our results suggest that continuous shearing motion results in the eruption of
the hot flux rope corresponding to flare eruption, 
whereas the anti-shearing motion causes the radiative condensation of the 
prominence formation.
The eruptive mechanism coupling with flux emergence is also the subject to investigate  
\citep{ChenShibata2000ApJ,Shiota2005ApJ,KanekoYokoyama2014ApJ}.
The results of our simulation with the good mechanical balance can be directly adopted 
as the initial condition for these eruptive studies. 
The model including the dense materials of the prominence may reveal 
the more detailed structure of coronal mass ejection 
such as the dense core and its evolution. 

\begin{figure}[htbp]
  \begin{tabular}{cc}
    \begin{minipage}{0.5\hsize}
      \includegraphics[scale=0.4,bb=0 0 566 425]{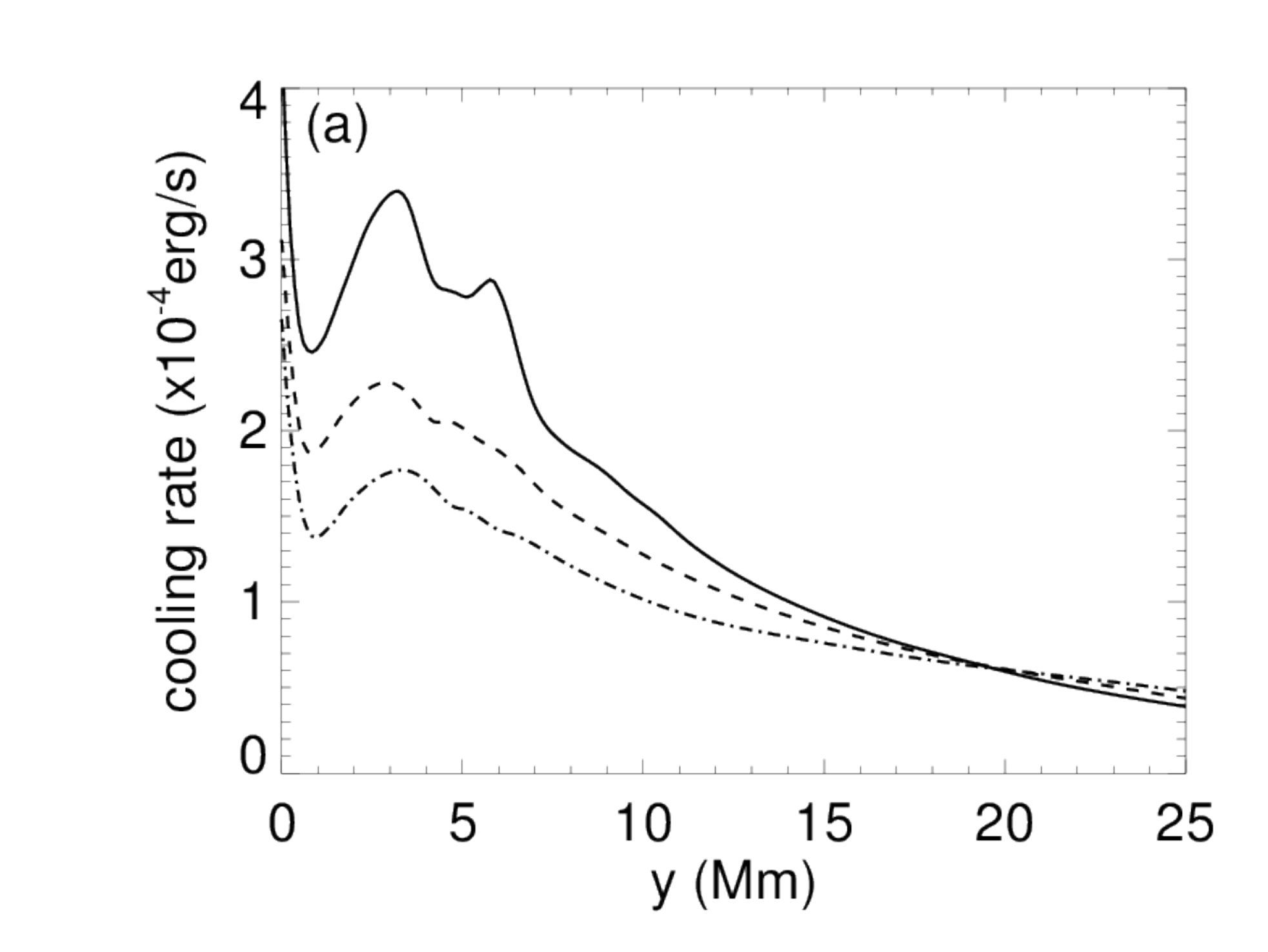}
    \end{minipage}
    \begin{minipage}{0.5\hsize}
      \includegraphics[scale=0.4,bb=0 0 566 425]{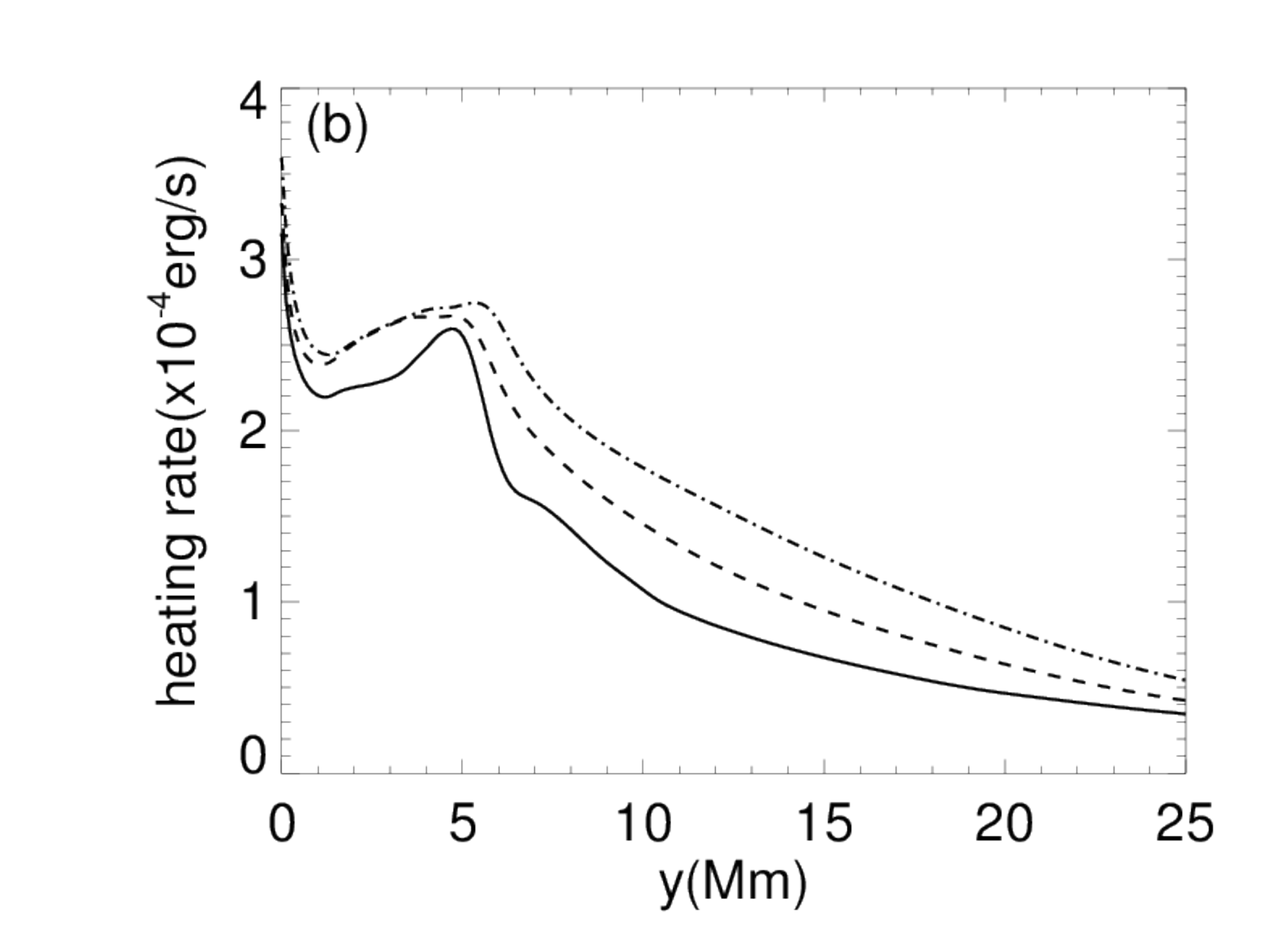}
    \end{minipage}
  \end{tabular}
  \caption{Cooling and heating rates along the $y$-axis at time=$500$s 
    in Fig.\ref{snap_c_llr}. Solid, dashed, and dot-dashed lines represent cases M3, M4 and M5,
    respectively.}
  \label{clht}
\end{figure}

\begin{figure}[htbp]
  \begin{center}
    \includegraphics[scale=0.5,bb=0 0 566 425]{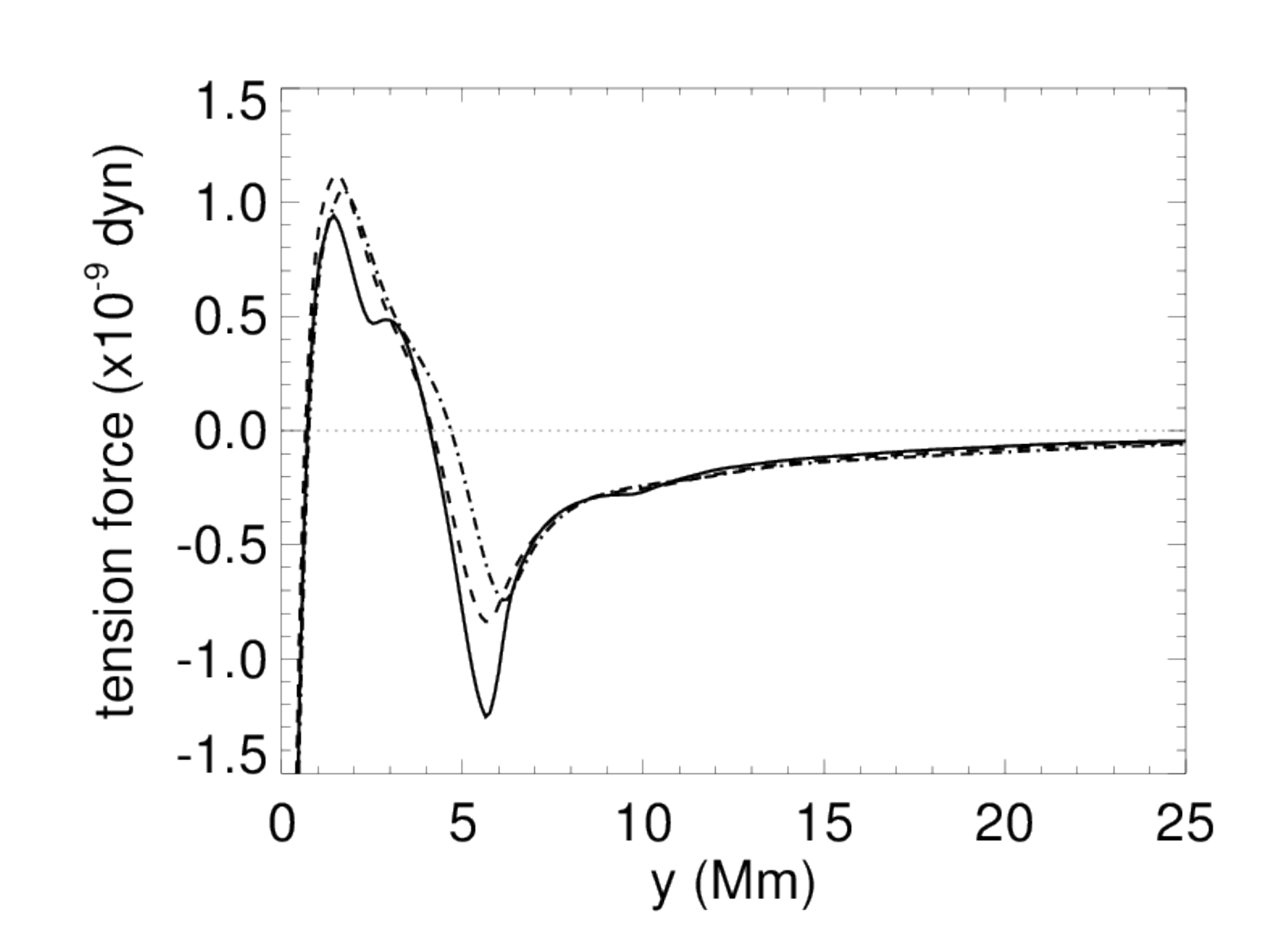}
  \end{center}
  \caption{The vertical magnetic tension force along the $y$-axis
    at time=$500$s. Solid, dashed, and dot-dashed lines represent M3, M4, and M5,
    respectively.}
  \label{tens}
\end{figure}

\begin{figure}[htbp]
  \begin{tabular}{cc}
    \begin{minipage}{0.5\hsize}
      \includegraphics[scale=0.4,bb=0 0 566 425]{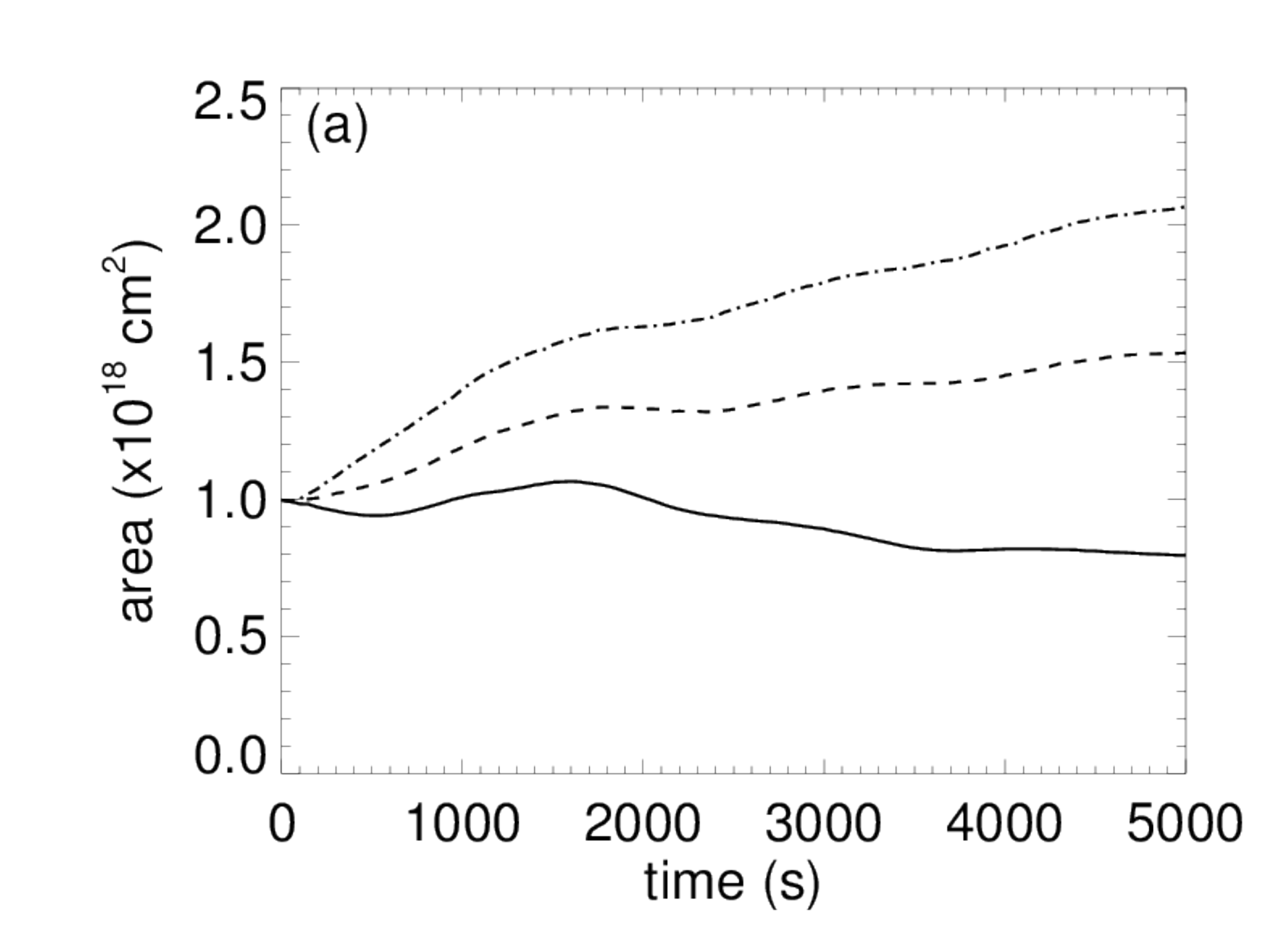}
    \end{minipage}
    \begin{minipage}{0.5\hsize}
      \includegraphics[scale=0.4,bb=0 0 566 425]{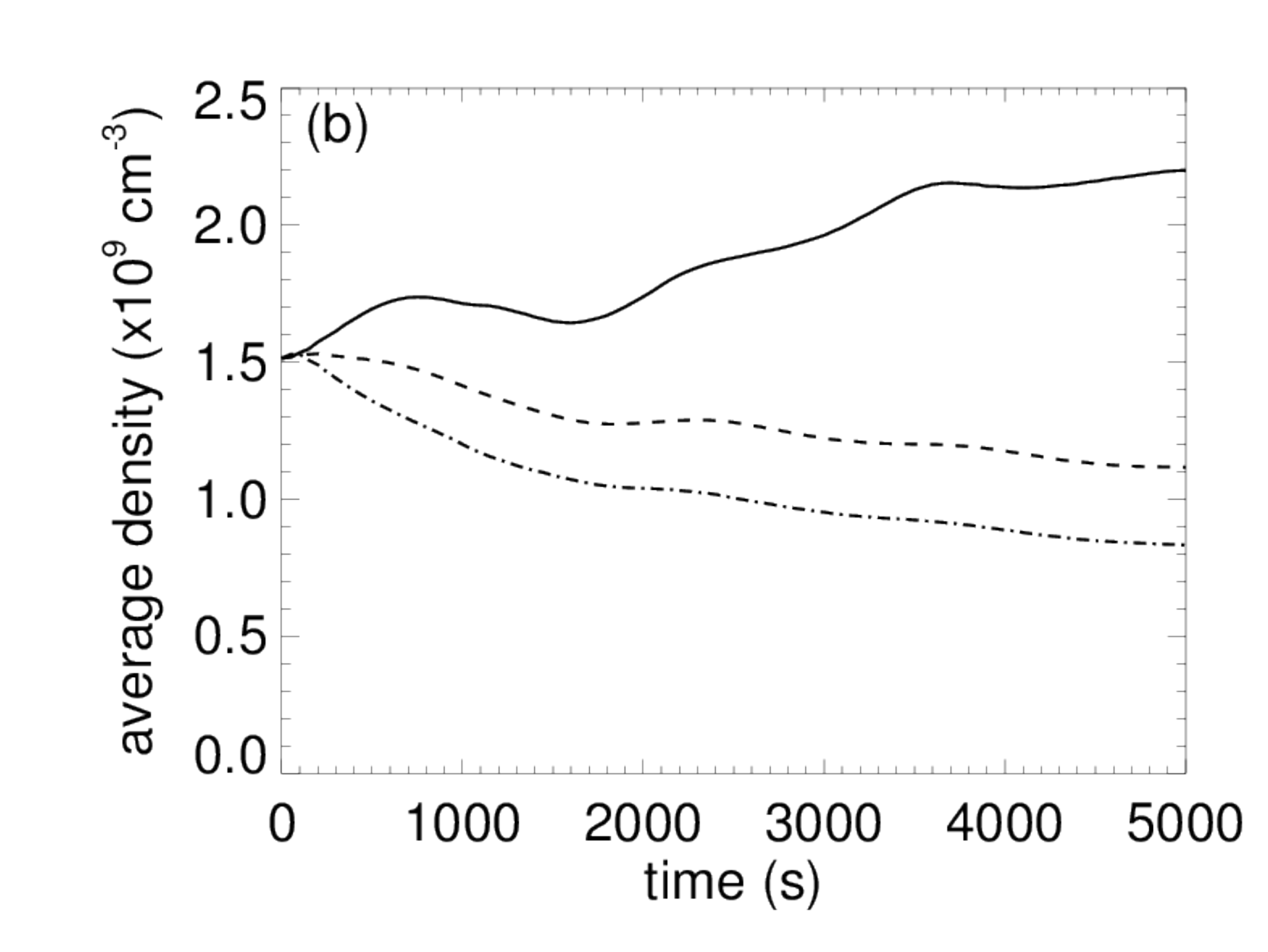}
    \end{minipage}
  \end{tabular}
  \caption{Time evolution of area and average density inside the dash-dotted line 
    in Fig.\ref{snap_c_llr}. Solid, dashed, and dash-dotted lines correspond to
    M3, M4, and M5, respectively}
  \label{dena}
\end{figure}

\begin{figure}[htbp]
  \begin{tabular}{cc}
    \begin{minipage}{0.5\hsize}
      \includegraphics[scale=0.4,bb=0 0 566 425]{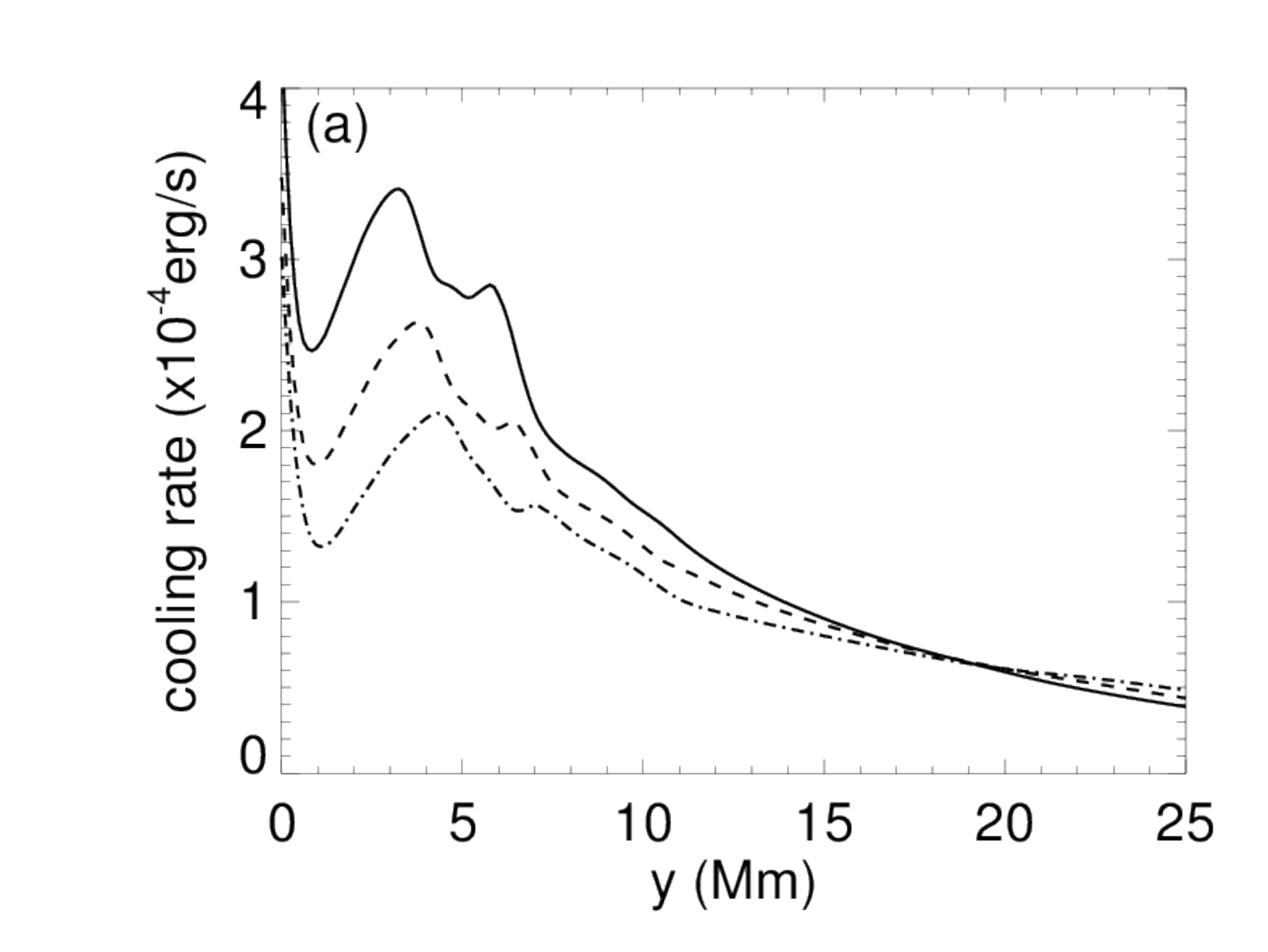}
    \end{minipage}
    \begin{minipage}{0.5\hsize}
      \includegraphics[scale=0.4,bb=0 0 566 425]{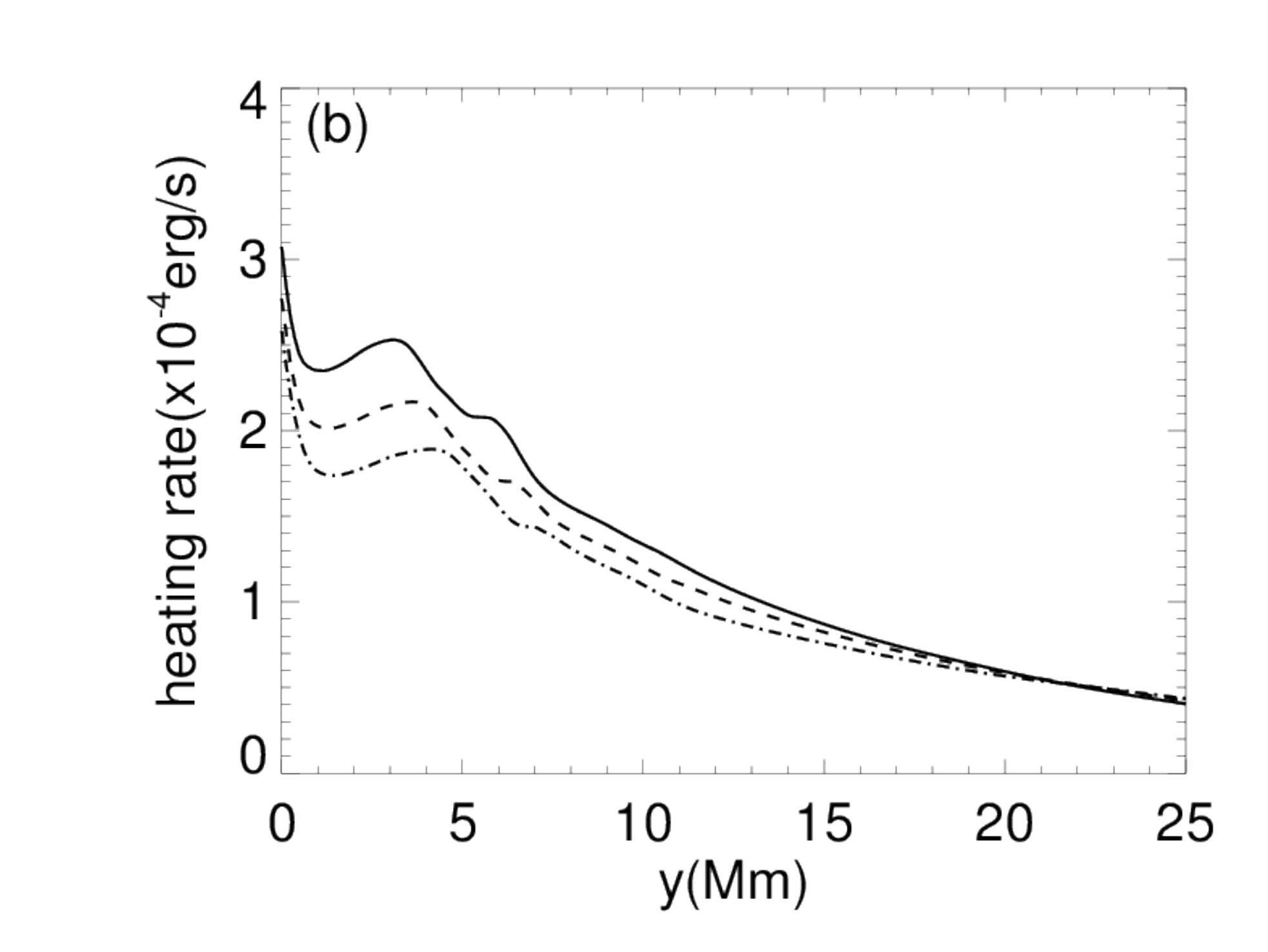}
    \end{minipage}
  \end{tabular}
  \caption{Cooling and heating rates along the $y$-axis at time=$500$s 
    in Fig.\ref{snap_c_llr}. Solid, dashed and dot-dashed lines represent cases D3, D4 and D5,
    respectively.}
  \label{clht_ro}
\end{figure}

\subsection{Comparison with observation}
We compare the simulated process of in-situ condensation with the observation
by \citet{Berger2012} by synthetic emission through the extreme ultraviolet
(EUV) filters of the Solar Dynamics Observatory Atmospheric Imaging Assembly (SDO/AIA).
The emission of a certain wavelength channel $i$ is expressed as
\begin{equation}
  D_{i}=\int n_{e}^{2} K_{i}(T) dl, 
  \label{em}
\end{equation}
where $D_{i}$, $n_{e}$, $K_{i}(T)$, and $l$ 
represent the pixel response to the photon flux, electron number density, 
temperature response function of the AIA filters, and the distance along the line-of-site
(LOS), respectively.
The temperature response functions are obtained from 
aia\_get\_response.pro in the Solar Software Library.
To compute the synthetic emission, we choose the $z$-axis as the LOS.
In our 2.5-dimensional simulation, the physical variables are assumed to be constant
along $z$-axis; hence, the integration along the LOS is altered 
to the multiplication of an arbitrary constant $l_{z}$ as follows,
\begin{equation}
  D_{i}=n_{e}^{2} K_{i}(T) l_{z} .
  \label{em_z}
\end{equation}
Figure \ref{synz} is the time evolution of the synthetic emission of each EUV filter.
Note that Fig. \ref{synz} covers the time only before the temperature reaches the cutoff.
Figure \ref{synz} clearly shows that the areas of strong emission progressively shift 
from the 193$\mrm{\AA}$ channel, through 171$\mrm{\AA}$ channel, 
to 131$\mrm{\AA}$ channel both in time and altitude. 
The emission shift was claimed to be evidence of in-situ condensation 
in \citet{Berger2012}, and our simulation has validated their argument.
The shift from the 211$\mrm{\AA}$ channel to the 193$\mrm{\AA}$ channel is not so clear
because the initial coronal temperature is set to 1MK.
The shift in altitude in Fig. \ref{synz} results from the descent of 
the flux rope and high density plasma.
It can also be realized by some other mechanisms 
such as magnetic reconnection in the weighted dip \citep{Liu2012ApJ,KeppensXia2014ApJ},
or cross-field slippage \citep{Low_a_2012ApJ,Low_b_2012ApJ}.
These mechanisms are not realized in our simulations because we set the
cutoff temperature such that the simulated prominence density can not 
be sufficiently large.

Figure \ref{romax} shows the relationship between the temperature and the maximum density 
of the condensations obtained from our results. 
The lower the condensation temperature, the higher the density of the condensates. 
The resultant density is independent of the heating model.
The temperature and density of the prominence are related by the following scaling formula, 
\begin{equation}
  \left( \frac{n}{2\times 10^{9}~\mrm{cm^{-3}}}\right)
  =85 \left( \frac{T}{10^{4}~\mrm{K}} \right)^{-1}.
  \label{scaling}
\end{equation}
The scaling relationship (Eq. (\ref{scaling})) is plotted as a
solid line in Fig. \ref{romax}.
This formula indicates that the dense condensation results from
re-stratification along the magnetic field lines with temperature $T_\mrm{c}$. 
Because the mass is conserved along the field lines and 
confined to the length scale of the scale height $k_{B}T_\mrm{c}/(mg_\mrm{cor})$, 
the condensate density is inversely proportional to temperature.
From Eq. (\ref{scaling}), we concluded that our model 
can reproduce the observed prominence density, which is 10--100 times larger than 
the surrounding coronal density at typical temperatures (below $10^{4}~\mrm{K}$). 

The direct simulation without the temperature cutoff is necessary to verify 
the realistic prominence density and compare DEM with observation. 
Including the chromosphere in our model is also future work. 
Due to the thermal conduction from the corona
to chromosphere, the background heating rate must be larger than the cooling rate.
When the initial coronal density is lower, the magnitude of the initial thermal imbalance is lower,
and the present model without the initial thermal imbalance will be validated.
It is the next step for our model to be extended to three dimensions. 
Self-consistent three-dimensional MHD simulation including coronal thermal 
conduction is still a challenging issue. The comparison with observations including the
LOS effect \citep{LunaKarpen2012ApJ,XiaKeppens2014}
must contribute to the further understanding of the prominence formation.

\begin{figure}[htbp]
  \begin{center}
    \includegraphics[scale=0.5,bb=0 0 708 708]{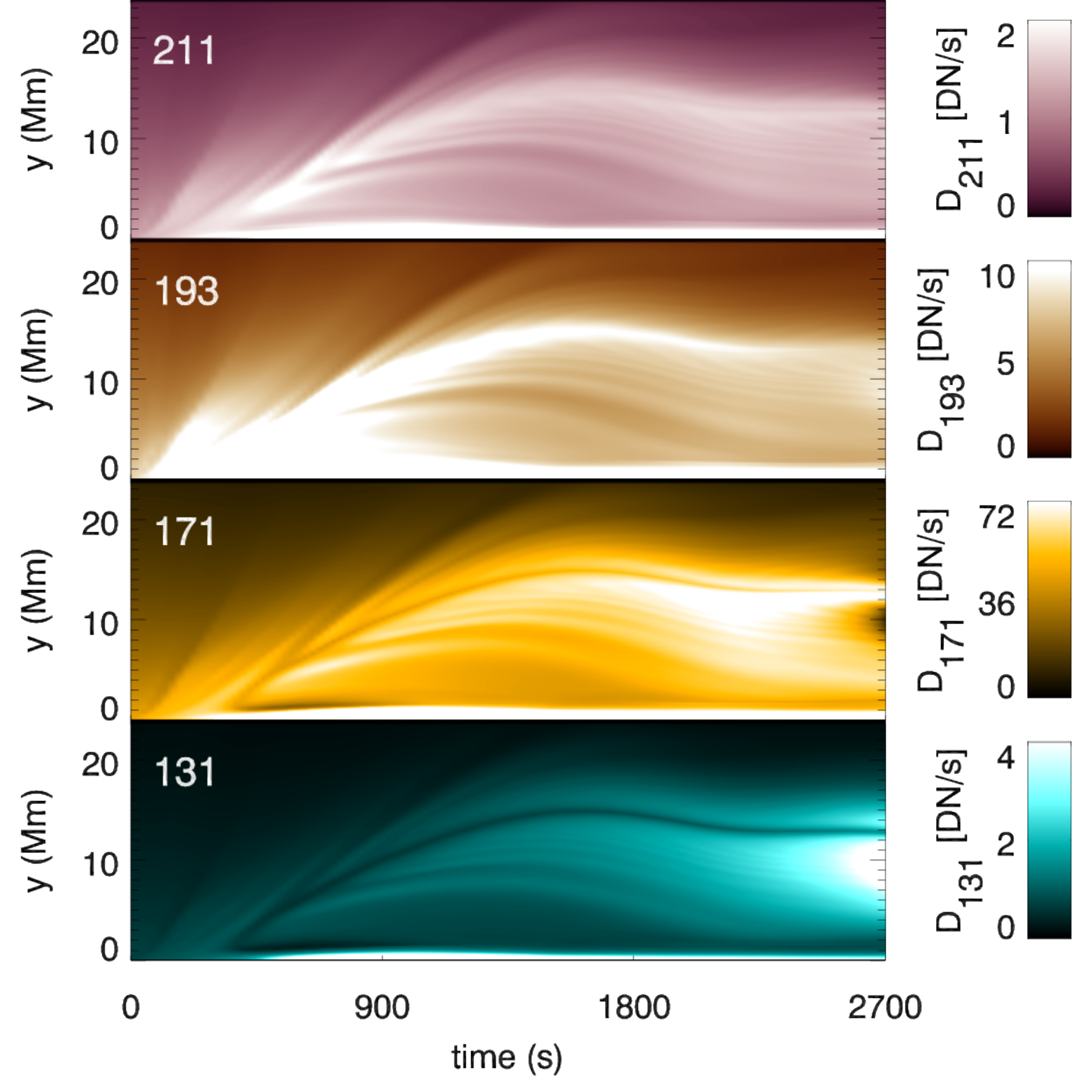}
    \caption{Time evolution of synthetic EUV emission 
      through the 211$\mrm{\AA}$, 193$\mrm{\AA}$, 171$\mrm{\AA}$, 
      and 131$\mrm{\AA}$ channels of AIA.
      $\log T_{\mrm{peak}}$ = 6.2, 6.1, 5.8, and 5.6, respectively, where 
      $T_{\mrm{peak}}$ is the peak temperature of each channel.
      $l_{z}$ in Eq. (\ref{em_z}) is set to 10 Mm.
      The emission is averaged in the area of $x<3\mrm{Mm}$.}
    \label{synz}
  \end{center}
\end{figure}

\begin{figure}
  \begin{center}
    \includegraphics[scale=0.8,bb=0 0 425 340]{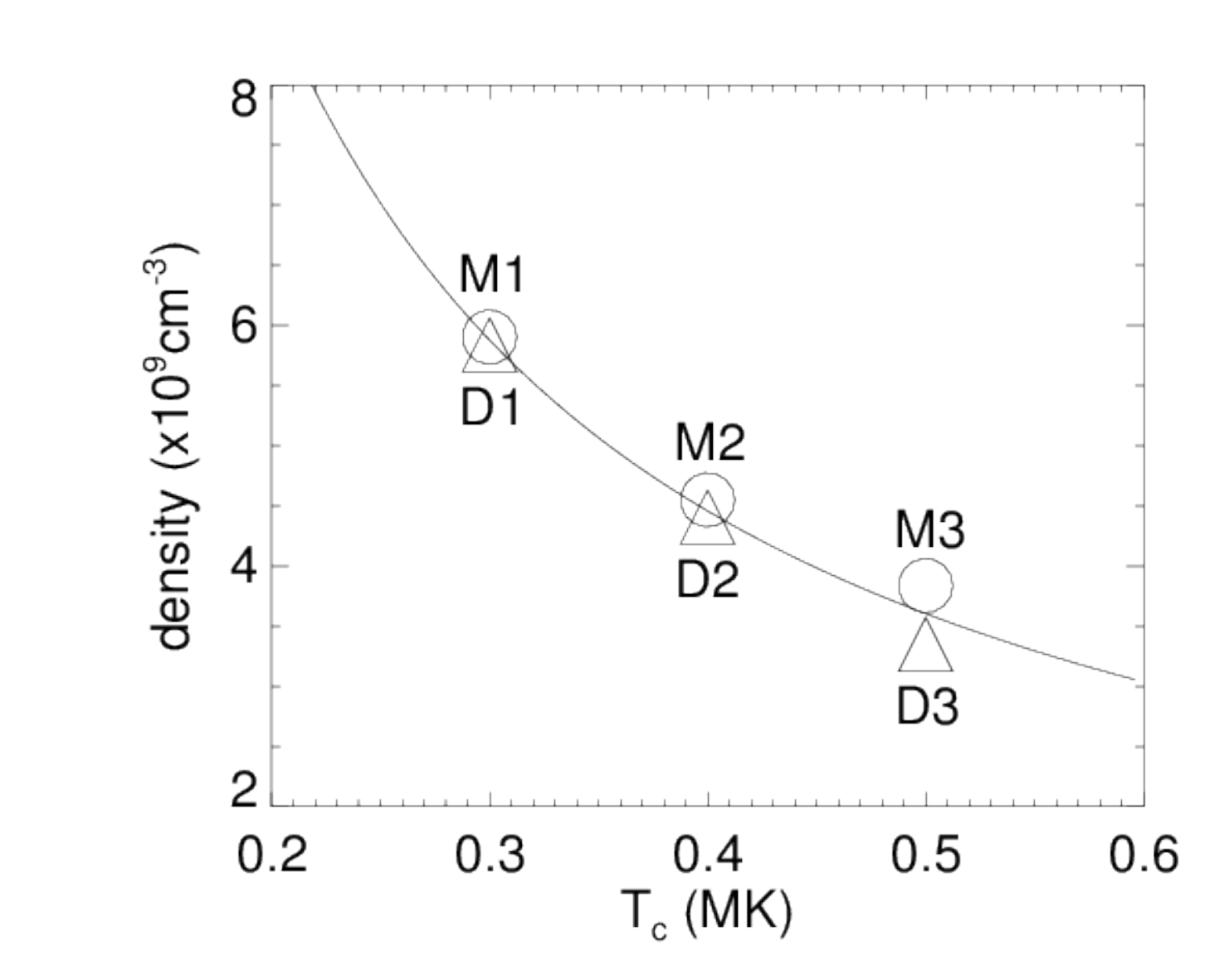}
    \caption{Relationship between $\mrm{T_{c}}$ and maximum density.
      The modeled cases are indicated by letters.
      The solid line is the result of applying the scaling formula Eq. (\ref{scaling}).}
    \label{romax}
  \end{center}
\end{figure}

\section{Conclusion}
We proposed a new in-situ formation model of inverse-polarity prominence.
The model was demonstrated in 2.5-dimensional MHD simulations, 
involving thermal conduction and radiative cooling.
Either flux rope formation by anti-shearing motion or 
a heating model proportional to the local density is necessary
for a in-situ prominence formation. 
From the results of a parameter study on cutoff temperature, 
we derived an empirical scaling formula that related the density 
of the prominences to their temperature.
Our study quantitatively demonstrated that our in-situ formation
model can reproduce the observed density of prominences.
The study also explained the observed intensity shift among multi-wavelength EUV emission
of in-situ condensation.

\acknowledgments
We are grateful for the helpful discussion with Dr. C. Xia and Prof. Dr. R. Keppens 
during my short stay in KU Leuven. This trip was supported by the Program
for Leading Graduate School, MEXT, Japan.
This work was supported by JSPS KAKENHI Grant Number 15H03640.
Numerical computations were conducted on a Cray XC30 supercomputer 
at the Center for Computational 
Astrophysics (CfCA) of the National Astronomical Observatory of Japan.
The authors would like to thank Enago (www.enago.jp) for the English language review.

\end{document}